\documentclass[twocolumn]{elsart3p}

\usepackage{times}
\usepackage{graphicx}
\usepackage{amssymb}
\usepackage{amsmath}

\newcommand{\degr}{$^{\circ}$}

\newcommand{\aap}{Astronomy \& Astrophysics}

\newcommand{\apj}{Astrophysical Journal}
\newcommand{\apjl}{Astrophysical Journal Letters}

\newcommand{\apss}{Ap\&SS}

\newcommand{\jgr}{J. Geophys. Res.}
\newcommand{\mnras}{Royal Astronomical Society, Monthly Notices}
\newcommand{\nat}{Nature}

\newcommand{\physrep}{Physics Reports}
\newcommand{\prd}{Phys. Rev. D}

\begin{document}

\begin{frontmatter}

\title{Search for Primordial Black Holes with SGARFACE}

\author[isu]{M. Schroedter\corauthref{cor}},
\corauth[cor]{Corresponding author.}
\ead{schroedt@iastate.edu}
\author[isu]{F. Krennrich},
\author[utah]{S. LeBohec},
\author[pennstate]{A. Falcone},
\author[paris]{S.~J. Fegan},
\author[paris]{D. Horan},
\author[whipple]{J. Kildea},
\author[argonne]{A.~W. Smith},
\author[nui]{J. Toner},
\author[whipple]{T.~C. Weekes}

\address[isu]{Dept. of Physics and Astronomy, Iowa State University, Ames, IA 50011-3160 }
\address[utah]{Dept. of Physics, University of Utah, Salt Lake City, UT 84112-0830 }
\address[pennstate]{Pennsylvania State University, 525 Davey Laboratory, University Park, PA 16802}
\address[paris]{Laboratoire Leprince-Ringuet, Ecole Polytechnique, CNRS/IN2P3, F-91128 Palaiseau, France}
\address[argonne]{Argonne National Lab, High Energy Physics, Argonne, IL 60439}
\address[nui]{School of Physics, National University of Ireland, Galway, Ireland}
\address[whipple]{Whipple Observatory, Harvard-Smithsonian Center for Astrophysics, P.O. Box 6369, Amado, AZ 85645-0097}
\begin{abstract}
The Short GAmma Ray Front Air Cherenkov Experiment (SGARFACE) uses the Whipple 10~m telescope to search for bursts of $\gamma$ rays. SGARFACE is sensitive to bursts with duration from a few ns to $\sim$20~$\mu$s and with $\gamma$-ray energy above 100~MeV. SGARFACE began operating in March 2003 and has collected 2.2 million events during an exposure time of 2267~hours. A search for bursts of $\gamma$ rays from explosions of primordial black holes (PBH) was carried out. A Hagedorn-type PBH explosion is predicted to be visible within 60~pc of Earth. Background events were caused by cosmic rays and by atmospheric phenomena and their rejection was accomplished to a large extent using the time-resolved images. No unambiguous detection of bursts of $\gamma$ rays could be made as the remaining background events mimic the expected shape and time development of bursts. Upper limits on the PBH explosion rate were derived from the SGARFACE data and are compared to previous and future experiments. We note that a future array of large wide-field air-Cherenkov telescopes equipped with a SGARFACE-like trigger would be able to operate background-free with a 20 to 30 times higher sensitivity for PBH explosions.
\end{abstract}

\end{frontmatter}
\section{Introduction \label{sec:introduction}}
SGARFACE is designed to detect short bursts of $\gamma$ rays with energies above 100 MeV using the air-Cherenkov technique \cite{Sgarface_Krennrich2000}. The design allows detection of bursts lasting up to $\sim$20 $\mu$s and has a directional accuracy of up to a few arc-minutes \cite{Sgarface_Lebohec2005}. Bursts of this type could be produced by the evaporation of primordial black holes (PBH), from emission associated with giant radio pulses from pulsars \cite{Lyutikov2008}, or by a high-energy component accompanying very short gamma-ray bursts.

Primordial black holes would have been the first objects formed by the gravitational collapse of inhomogeneities in the early universe \cite{Zeldovich1967,Hawking1971,Carr_Hawking1974}. Inhomogeneities may have resulted from density fluctuations \cite{Carr1975} present after inflation \cite{Gilbert1995}, the presence of cosmic strings \cite{Polnarev1991,MacGibbon_Brandenberger1998}, collapse of domain walls \cite{Rubin2001}, or from bubble collisions during a phase transition \cite{Hawking1982}. Consequently, even without a detection, limits on the PBH density provide important insight into many aspects of the physics of the early universe \cite{Carr2005}, including: cosmological phase transitions \cite{Kapusta2007} and references therein, structure formation \cite{Afshordi2003,Khlopov2005}, mass clustering \cite{Chisholm2006}, and magnetic monopoles \cite{Stojkovic2005}.

Three effects have been used to search for PBHs:

(1) Gravitational effects: Black holes of mass $\gtrsim 10^{-8}$ $M_\odot$ ($2\times10^{25}$ g) may be detectable by their gravitational effects in microlensing observations \cite{Paczynski1986,Alcock1996} or from mass clustering measured with the Ly$\alpha$ forest \cite{Afshordi2003}. A significant number of heavier PBHs may have formed binary systems \cite{Nakamura1997} and may be detected from the gravitational radiation emitted during coalescence. An upper limit on the BH density in the mass range from 0.2 to 1.0 $M_\odot$  has been reported by the LIGO collaboration \cite{Abbott2005}.

(2) Hawking radiation: Black holes of mass less than $10^{14}$~g may be detected by their emitted Hawking radiation \cite{Hawking1974}. Black holes of mass, $M$, emit real particles with a black-body energy spectrum at temperature $T = 1.06 (10^{13} \mathrm{g}/M)$ GeV \cite{MacGibbon_Webber1990}. As mass accretion by PBHs is thought to be negligible \cite{Custodio2002}, PBHs will evaporate completely in $10^{64}(M/M_\odot)^3$ years, where $M_\odot\approx 2\times10^{33}$g is the Solar mass. At the present time, black holes with an initial mass of $4.5 \times 10^{14}$~g \cite{MacGibbon1991} would be reaching the final phase of their evaporation which might be explosive. PBHs that have already evaporated, would have left an imprint on the cosmic particle backgrounds and set stringent limits on the PBH density; a discussion is given in Sect.~\ref{sec:PBH_searches}. Lastly, PBHs may also be detectable as transient gamma-ray sources in the solar system \cite{Page_Hawking1976,MacGibbon_Carr1991}.

(3) Accretion effects: The accretion onto light black holes may produce distinct observable radiation in the relatively dense environments of interstellar space, in binary systems, or near planets~\cite{Trofimenko1990}.

The types of particles emitted by Hawking radiation and their energy spectra are well understood below the QCD confinement temperature of $T_c \approx 175^{+5}_{-18}$ MeV \cite{Braun_Munzinger2004}. Above this temperature, it is unknown whether individual elementary particles will be emitted in the strong gravitational field, or if a phase transition of the surrounding vacuum occurs and a thermodynamic model of the emission is valid. If the standard particle physics model (SM) is valid at all temperatures, the emission spectra can be determined from the quark-gluon jet fragmentation functions convolved with the Hawking emission function \cite{MacGibbon_Webber1990,MacGibbon1991}. This would result in a slow evaporation at high temperatures: the final explosion would last about 1 s with $\gamma$-rays of energies above 400 GeV being emitted \cite{Halzen1991,Connaughton1998}. 22\% of the total energy would be emitted in $\gamma$ rays with a flux spectrum peaking at $\sim$ 100 MeV and falling off as $E^{-3}$ at higher energies \cite{MacGibbon1991}.

On the other hand, in thermodynamic models of the QCD phase transition point, the large number of hadronic resonances implies a large phase space and therefore the final evaporation becomes extremely rapid, resulting in an explosion. The most extreme thermodynamic model is that of \cite{Hagedorn1965,Hagedorn1970}, referred to as the Hagedorn model (HM), where the density of states, $\rho_s$, increases exponentially with mass: $\rho_s \propto m^{-5/2} e^{m/T_c}$. In this picture, the ground state of the vacuum contains elementary and composite particles, which are evaporated by the gravitational field of the PBH. During the final explosion, 10-30\% of the mass at $T_c$ is converted into gamma-rays between 100 MeV and 1 GeV; the explosion lasts $\sim$100~ns~\cite{Page_Hawking1976}.

Instead of completely evaporating, PBHs may leave behind Planck mass relics, see \cite{Carr2005} and references therein. Because of their small mass, Planck relics do not affect $\gamma$-ray emission, but may possibly contribute to dark matter \cite{MacGibbon1987,Barrau2004}. If extra dimensions exist below the currently measured scale of gravity, $\sim$~10~$\mu$m \cite{Ferrari2006}, black hole evaporation would release less detectable energy~\cite{Argyres1998,Nouicer2007}.

For SGARFACE, we consider the detection of PBHs via their final-stage $\gamma$-ray radiation in the Hagedorn model. The nominal burst is assumed to come from a black hole of mass $6.5\times 10^{13}$ g, corresponding to $T$ = 160 MeV. It is assumed that 20\% of the mass-energy, or $7.5\times 10^{45}$ eV, is emitted in the form of 1~GeV $\gamma$ rays. Realistically, the $\gamma$-ray spectrum would not be a $\delta$-function, but a product of the direct $\gamma$ emission with the decays from hadrons. It is straightforward to scale the upper limits on the PBH explosion rate presented in Section~\ref{sec:bursts} with the total burst energy and the average energy of $\gamma$ rays, see Section~\ref{sec:volume}.

\subsection{Summary of PBH Searches \label{sec:PBH_searches}}
Limits on the PBH explosion rate have been placed by direct searches for the final stage emission and require a particular particle physics model. Limits on the PBH density that have been derived from cosmic-particle background spectra require an assumed initial PBH mass spectrum. To compare the results, assumptions need to be made on PBH clustering in the galactic halo \cite{Halzen1991,Wright1996,Chisholm2006} and, in the case of charged particle backgrounds, the enhancement due to confinement by the Galactic magnetic field \cite{MacGibbon_Carr1991}.

\begin{figure}[!ht]
    \includegraphics[width=0.48 \textwidth,clip]{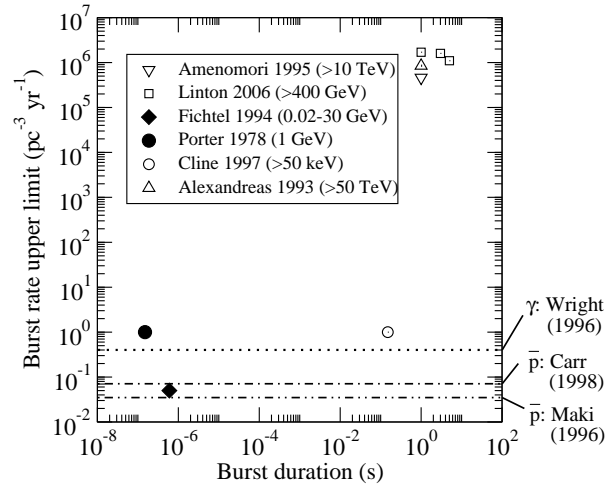}
\caption{Limits on the density of PBH explosions. Direct-detection limits are at 99\% CL, except for \cite{Cline1997}, and use $\Omega_m = 0.06$. The limits from particle background measurements have possibly large systematic uncertainties associated with them. }
  \label{fig:intro_limits}
\end{figure}

Air-Cherenkov telescopes (ACT), due to their large collection area, but small field of view of $\sim$4\degr, are well suited to search for individual explosions. The sensitivity to detect standard model PBH explosions is limited to distances of 0.4 pc \cite{Connaughton1998}, and to about 200 pc for the HM bursts. Upper limits on the explosion rate are shown in Fig.~\ref{fig:intro_limits}, where Hagedorn burst models have been assumed at time scales shorter than 1 $\mu$s and SM bursts at longer time scales. The first search for HM bursts was carried out by \cite{Porter_Weekes1977,Porter_Weekes1978,Porter_Weekes1979}, who arrived at an upper limit for bursts of 150 ns duration. The results of their two experiments were carefully reanalyzed using air-shower simulations to characterize the trigger threshold across the field of view. The revised value of their more sensitive, short-baseline, experiment is shown in Fig.~\ref{fig:intro_limits}.

The first search for SM bursts of duration between 0.01~s and 1~s and energy above 7~TeV was carried out by \cite{Fegan1978} with a pair of scintillation particle detectors separated by 250~km. Several groups, using single or closely spaced detectors, have set limits at other energies and duration \cite{Bhat1980_gamma,Nolan1990,Alexandreas1993,Amenomori1995,Connaughton1998,Linton2006}. Some of the limits shown in Fig.~\ref{fig:intro_limits} were standardized to the same SM burst by \cite{Halzen1991}. Satellites, though they have a small collection area, are useful in searching for PBH explosions because they are essentially background-free and have a large field of view. EGRET data was searched for multiple $\gamma$-ray tracks within its trigger window of 600$\pm$100~ns \cite{Fichtel1994}. Using BATSE data, a possible association of PBH explosions was made with some short (6-200~ms) X-ray bursts on the basis of the spectral evolution and spatial isotropy \cite{Cline1997}. 

The evaporation of PBHs over the lifetime of the universe would have left an imprint on the observed cosmic particle spectra: $\gamma$, $e^{\pm}$, $p$, $\bar{p}$, and $\nu$.%%%%
\footnote{We use the following constants: $H_0$=71~km/s/kpc, $13.6\times10^9$~yr as the age of the universe, and a critical density of $1\times10^{-29}$~g/cm$3$.}
%%%%%%
The underlying assumption in placing limits on the PBH density using particle backgrounds is that the initial PBH mass distribution arose from scale-invariant density fluctuations \cite{Page_Hawking1976}. Using the measured anisotropy in the EGRET $\gamma$-ray background above 100~MeV, \cite{Wright1996} determined the rate of PBH explosions to be less than 0.07-0.42~pc$^{-3}$yr$^{-1}$, the confidence level of this upper limit is not given. Of the charged particles, $\bar{p}$ are the most effective in setting limits on the PBH density \cite{MacGibbon_Carr1991,Carr_MacGibbon1998}, however additional uncertainties on the leakage time out of the Galaxy are introduced. The limit on the PBH explosion rate from the $\bar{p}$ data is (0.011-0.067)~pc$^{-3}$~yr$^{-1}$, where $\bar{p}$ are produced by jet-fragmentation. Here, the systematic error due to uncertainties in the halo enhancement may be as large as a factor of two \cite{Carr_MacGibbon1998}. The most recent and stringent upper limit comes from \cite{Maki1996}, who use a 3D Monte Carlo code to model the $\bar{p}$ propagation in the local Galactic environment and compare it to the BESS 1995 $\bar{p}$ data \cite{Yoshimura1995}. We scale their result to the 99\% CL as $3.4\times 10^{-2}$~pc$^{-3}$~yr$^{-1}$, assuming Poisson statistics. Another method to detect the evaporation of PBH relies on the neutrino background. The flux emitted in neutrinos is at least 2 to 3 times larger than the $\gamma$-ray flux~\cite{Carr1976}, but so far only the low-energy $\nu$-background has been used to constrain the PBH mass-distribution~\cite{Bugaev2002}.

Another idea to detect PBH explosions was put forward by \cite{Rees1977}, who suggested that a HM burst produces a sphere of simultaneously emitted charged particles, but see the arguments by~\cite{Heckler1997} and~\cite{MacGibbon2008}. These generate radio and optical emission by interaction with the ambient magnetic field. Searches for optical pulses have been carried out by \cite{Porter_Weekes1977_optical,Porter_Weekes1978} resulting in an upper limit of 0.3~pc$^{-3}$~yr$^{-1}$ at the 99\% CL. A dedicated search for coincident optical flashes between two optical telescopes with duration between 1 and 30~ms by \cite{Bhat1980_optical} set an upper limit at $0.028$~pc$^{-3}$~yr$^{-1}$.  Searches for radio pulses from PBH evaporation are summarized in \cite{Halzen1991} ; the best limit being $<7\times10^{-10}$~pc$^{-3}$~yr$^{-1}$. However, as the physical interaction model of \cite{Rees1977} is highly speculative, these upper limits should not be regarded as definite.

%%%%%%%%%%%%%%%%%%%%%%%%%%%%%%%%%%%%%%%%%%%%%%%%%%%%%%%%%%%%%%%%%%%%%%%%
%
%
%
%
%
\subsection{Description of SGARFACE}
SGARFACE~\cite{Sgarface_Lebohec2005} is now operating as part of the Whipple 10~m telescope, an imaging air-Cherenkov telescope (ACT) located in southern Arizona \cite{Crab_Nebula_Weekes89,Whipple_Kildea07}. The Whipple 10~m telescope is located at 31.6804\degr\ latitude, 110.8790\degr\ W longitude, and 2312~m a.s.l. 

The camera of the Whipple 10~m telescope consists of 379 close-packed PMTs covering a 2.4\degr\ field of view. The PMT signals for SGARFACE are split off the 10~m signal cables via passive couplers. To reduce cost and complexity of the SGARFACE system, clusters of seven nearest-neighbor pixels are summed into 55 channels, each viewing $\sim$0.36\degr\ of the sky. This pixelation is sufficient to image the extended Cherenkov images produced by bursts of $\gamma$ rays. The channels are sampled by flash analog-to-digital (FADC) converters at an interval of 20 ns and with memory depth of 35.04 $\mu$s (1752 samples). A trigger, described below, causes the FADC buffers to be stored to disk. The FADC input was designed with a time-constant of about 25 ns to preserve the accumulated charge between samples. SGARFACE achieves background rejection of cosmic rays and atmospheric phenomena by the time-resolved imaging of the Cherenkov wavefronts.

\begin{figure}[!ht]
    \includegraphics[width=0.4 \textwidth,clip ]{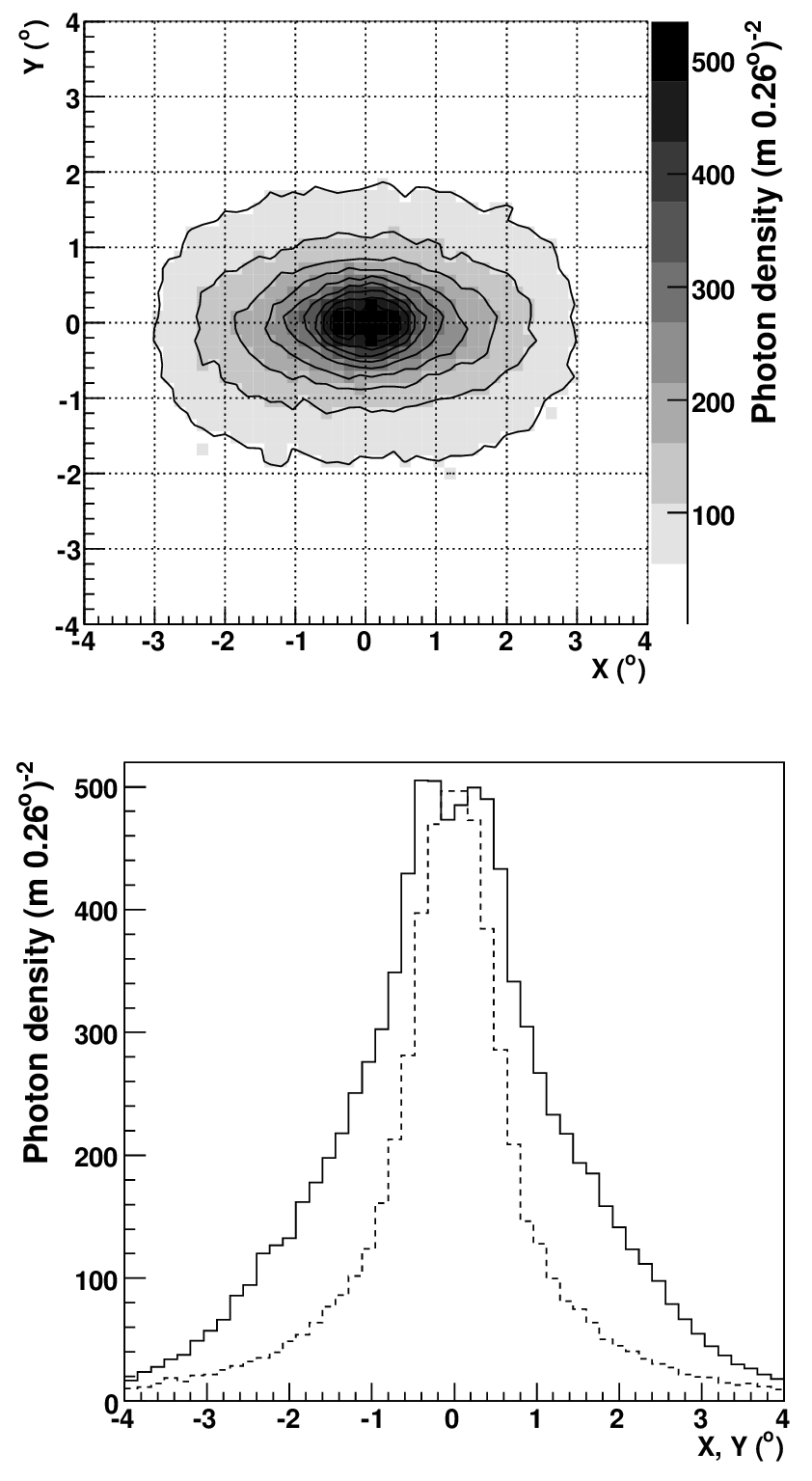}
  \caption{Angular distribution of Cherenkov light received at 2300~m altitude from a plane wave front of 1~GeV $\gamma$ rays at zenith with fluence of 4~$\gamma$/m$^2$. The x-axis is perpendicular to the Earth's magnetic field. The \emph{solid} line shows the photon distributions along the x-axis; the \emph{dashed} line along the y-axis.  }
  \label{fig:Cherenkov_distribution}
\end{figure}

\begin{figure}[!ht]
    \includegraphics[width=0.45 \textwidth,clip]{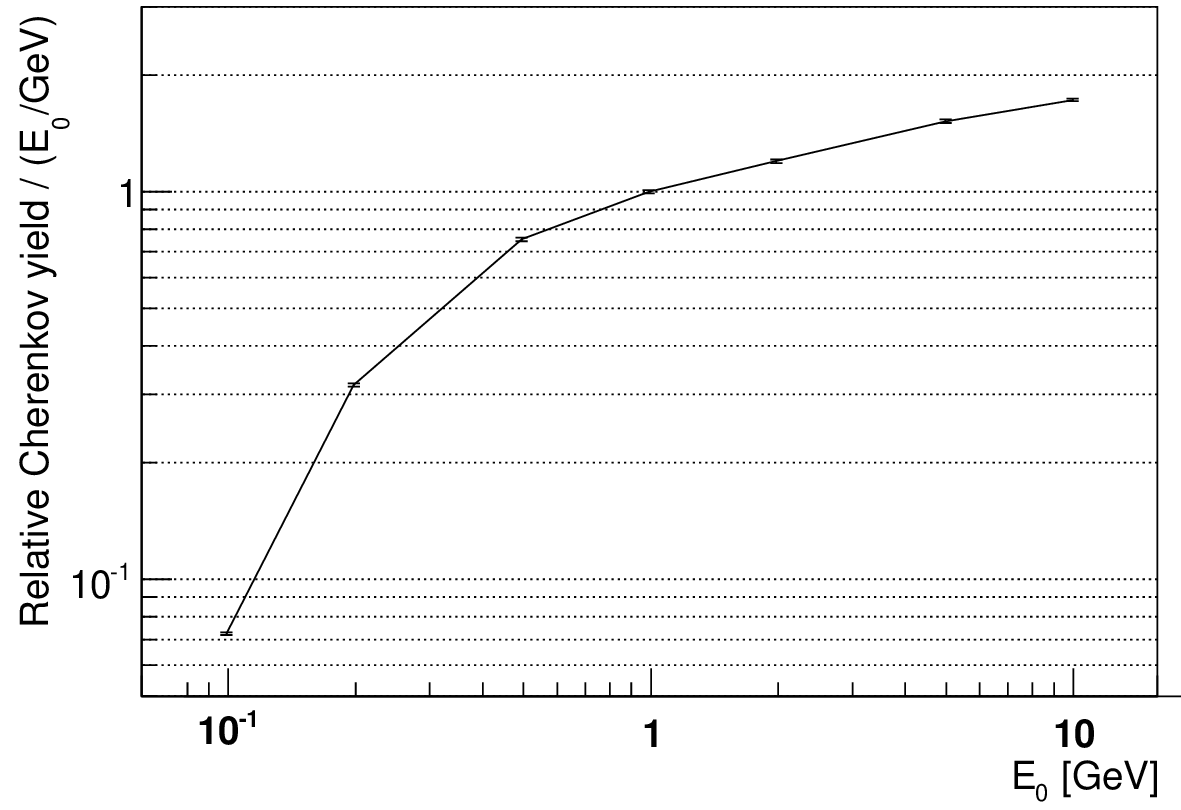}
  \caption{Variation of the Cherenkov photon yield divided by the primary energy, $E_0$, versus $E_0$, corresponding to the yield of Cherenkov photons for a burst with fixed total energy. The yield has been normalized relative to $E_0$ = 1~GeV. The yield decreases as the primary $\gamma$-ray energy approaches the threshold for Cherenkov emission. }
  \label{fig:sensitivity_energy}
\end{figure}

The air-Cherenkov technique has a detection area of $\sim 1\times 10^5$ m$^2$ for single TeV $\gamma$-ray primaries~\cite{Mohanty98}. At sub-GeV energies, the Cherenkov emission from one $\gamma$ ray is not detectable, but a wavefront of near-simultaneous sub-GeV $\gamma$-rays produces a detectable signal~\cite{Sgarface_Krennrich2000}. The time dispersion of Cherenkov photons from an instantaneous burst is 10~ns (35~ns) on-axis (3\degr\ off-axis) due to the shower geometry, while a 6.5~ns time-spread is introduced by the geometry of the 10~m telescope. The angular distribution of Cherenkov photons of these low-energy $\gamma$-ray wavefronts is smooth and broad, as shown in Fig.~\ref{fig:Cherenkov_distribution} for a burst of 1 GeV $\gamma$ rays at zenith. 
The widths of the angular distributions, shown in the lower part of Fig.~\ref{fig:Cherenkov_distribution}, are a combination of Compton and Coulomb scattering, and the geomagnetic bending of $e^{\pm}$. Taking into account the number of radiation lengths available for electrons until they reach the threshold for Cherenkov emission in the atmosphere, the total multiple Coulomb scattering angle is roughly proportional to $p^{-1/2}$, where $p$ is the initial electron momentum~\cite{PDG}. Similarly, the width of the photon angular distribution arising from Compton scattering is proportional to $p^{-1/2}$, where $p$ is the  photon momentum ~\cite[pg. 354-358]{LandauQED}. As the primary $\gamma$-ray energy decreases towards the Cherenkov threshold, relatively less Cherenkov radiation is produced from the total available energy; the threshold for $e^{\pm}$ is 80~MeV at 20~km altitude, the height of shower maximum for 100~MeV $\gamma$ rays. Therefore, the Cherenkov-light density on the ground depends strongly on the primary energy. For a fixed total energy, Figure~\ref{fig:sensitivity_energy} shows the Cherenkov photon yield with primary energy relative to 1 GeV $\gamma$ rays. 

The SGARFACE trigger consists of two systems~\cite{Sgarface_Lebohec2005}. A multi-time scale discriminator (MTSD) and a pattern-sensitive unit. The MTSD integrates the signal and triggers on six time scales: 60~ns, 180~ns, 540~ns, 1620~ns, 4,860~ns, and 14,580~ns. The trigger threshold requirement is tested at three equally spaced time intervals during the trigger window, requiring pulses to be at least as long as the trigger window. The thresholds were set conservatively above the night-sky background and occasionally adjusted to reflect changes in the telescope throughput. 

The pattern-sensitive coincidence unit determines if a sufficient number of nearest neighbor channels have triggered the MTSD. The majority of observation time was taken with a 7-fold nearest-neighbor coincidence requirement. Once an event triggers, the 55 FADC stacks are read out after a 16.28~$\mu$s delay. This data is stored together with information on telescope pointing, PMT voltage settings, and trigger information.

%
%%%%%%%%%%%%%%%%%%%%%%%%%%%%%%%%%%%%%%%%%%%%%%%%%%%%%%
%
%
%
\section{Data Analysis}
\subsection{Overview}
\begin{figure}[!ht]
    \includegraphics[width=0.45 \textwidth,clip]{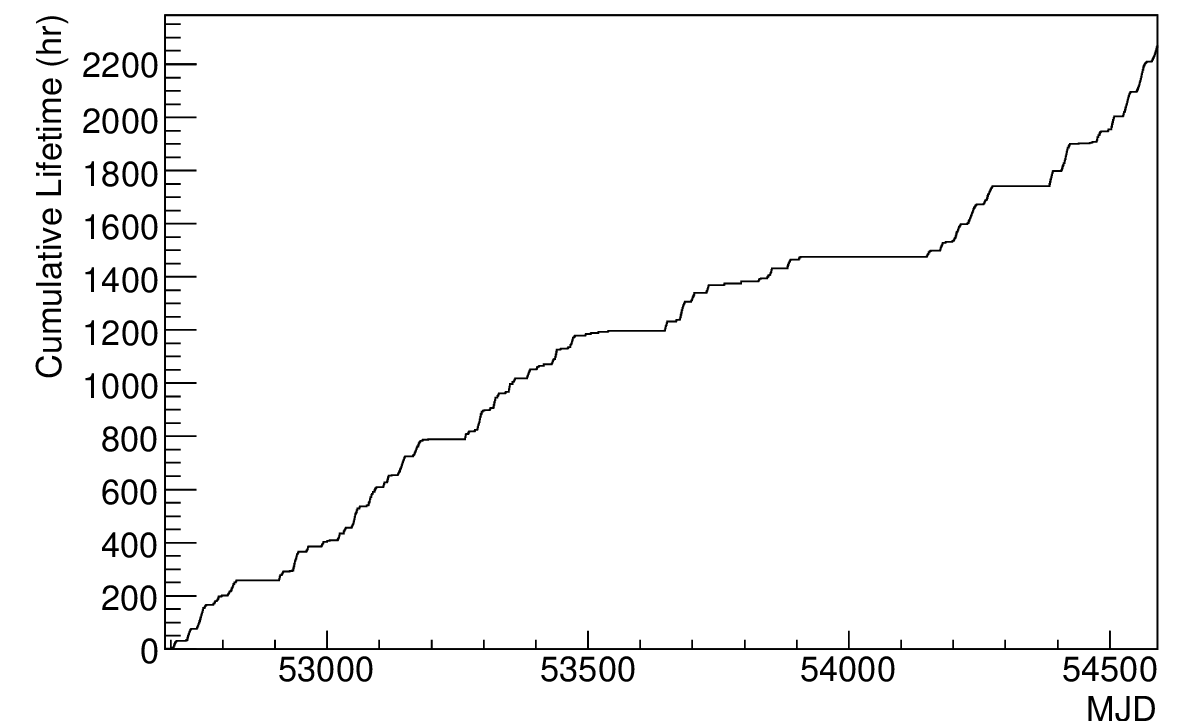}
  \caption{Lifetime accumulation occurs when the following conditions were met: trigger rate greater than 3 events/min, telescope pointing above 20\degr\ in elevation, telescope pointing varied by less than 4\degr\ root-mean square during the one minute interval. The last condition is not met when the telescope is being slewed to a new source.}
  \label{fig:cumulativeLTime}
\end{figure}

Since March 2003, SGARFACE has recorded about 2.2 million events during routine observations of the Whipple 10~m telescope. Observations with the 10~m telescope are usually only carried out when the weather conditions are judged acceptable for TeV astronomy observations. The mean data rate was 0.18~Hz. The dead-time of the data-acquisition (DAQ) system is 0.316 sec per event, and comprised 6\% of the total operational time. The lifetime accumulation is shown in Fig.~\ref{fig:cumulativeLTime}.

\begin{figure}[!ht]
    \includegraphics[width=0.48 \textwidth,clip]{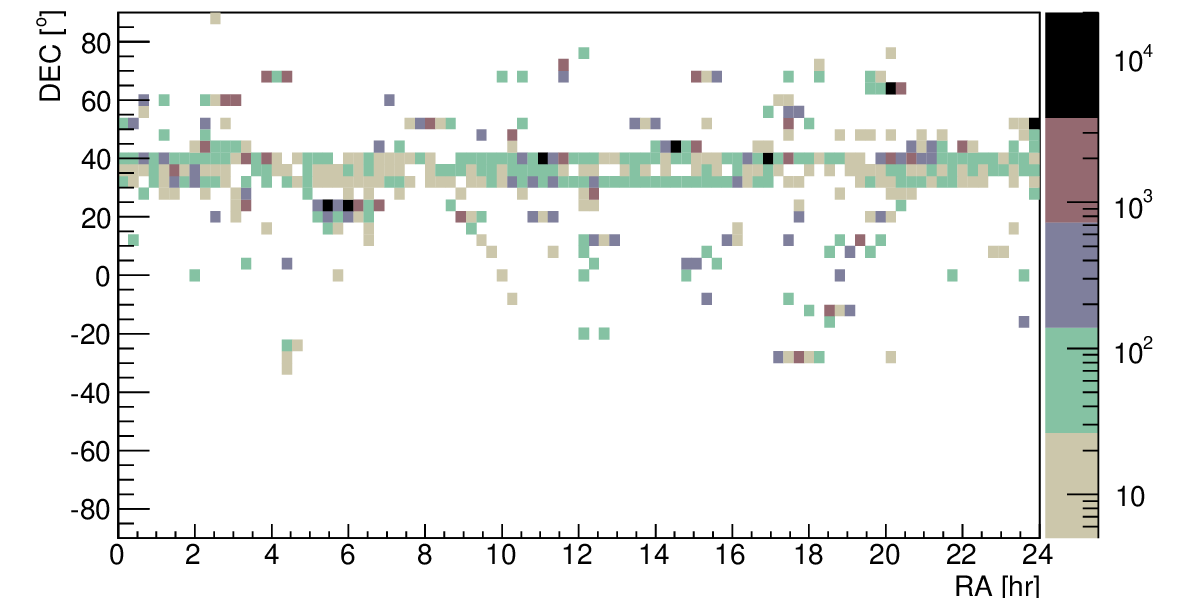}
    \includegraphics[width=0.48 \textwidth,clip]{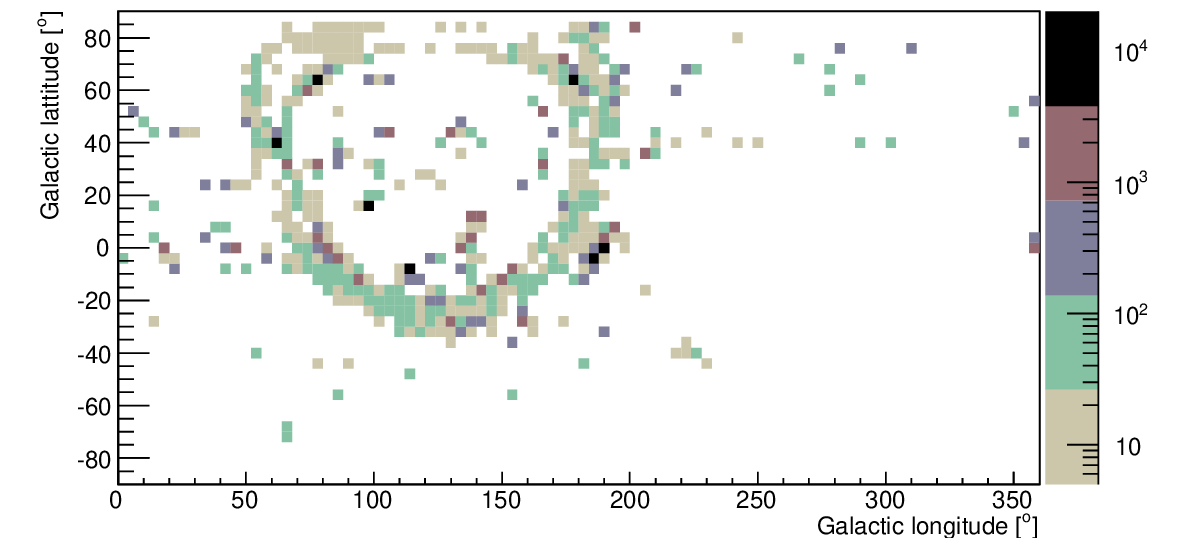}
    \includegraphics[width=0.48 \textwidth,clip]{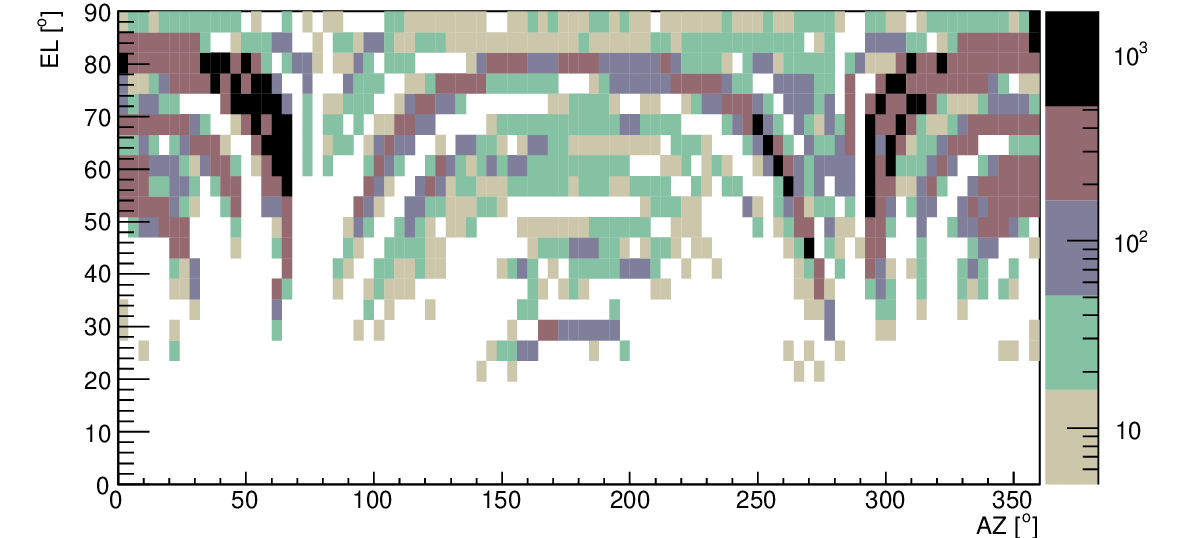}
\caption{Exposure map (minutes) in equatorial (\emph{top}), galactic (\emph{middle}), and horizon (\emph{bottom}) coordinates. For visual clarity, bins are suppressed with less than 5 min exposure. Some prominent points of observation are the Crab Nebula (RA 05h34m31.9s DEC+22\degr00m52.1s, J2000), Mrk~421 (RA 11h04m27.31s DEC+38\degr12m31.8s), and the Galactic Center (RA 17h45m40.0s, DEC-29\degr00m28.1s). The Galactic Center is near 360\degr\ longitude }
\label{fig:exposure_maps}
\end{figure}

The exposure of the sky during the six years of operation is shown in Figures~\ref{fig:exposure_maps} in equatorial, galactic, and horizon coordinates, respectively. Long exposures at prominent TeV gamma-ray sources are at the Crab Nebula, Mrk~421, Mrk~501, the Galactic Center, and H~1426+428. Near each source is an exposure offset in Right Ascension (RA) by $\pm$30 min that arises from Whipple OFF-source observations. The horizontal bands in the equatorial exposure map in Fig.~\ref{fig:exposure_maps} stem from the nightly 10~minute zenith runs when the telescope is pointed at or near the zenith.

\subsection{Event Analysis \label{sec:analysis} }
For each event, the topology and time development of the images is parameterized on five time scales. As in the hardware algorithm, the analysis sums the FADC traces with a top-hat filter function of width 1, 3, 9, 27, and 81 samples. The analysis then proceeds on each time scale as follows:

Fluctuations about the pedestal value are produced by night-sky background fluctuations, electronic noise from telescope-drive motors, and intrinsic electronics noise of the DAQ. In the analysis, the mean pedestal value is subtracted from the trace. The standard deviation of the distribution of the fluctuations, referred to as RMS in the following, is used to set the threshold that channels must cross before a particular sample is accepted as part of the image. Details of the pedestal stability are given in Sect.\ref{sec:hardware_stability}. The triggering sample in the FADC stack is located at least 814 samples before the end of the trace. Therefore, the pedestal of each channel is determined from the first 750 samples with a value less than 40~dc multiplied by the width of the top-hat filter. The longest trigger time scale of three times 243 samples, or 14.58~$\mu$s, is not used in the analysis because the pedestal value cannot be estimated accurately from the initial signal-free portion of the FADC trace. The loss in fluence sensitivity by excluding this time scale is minimal.

Next, FADC samples that contain little or no signal are suppressed followed by a moment-analysis to characterize the image shapes. The cleaning procedure proceeds in two steps: 1) Any re-binned samples with signal in excess of three times the pedestal RMS value are retained. 2) Samples adjacent in time and space are retained if their signal is greater than twice their pedestal RMS value. All other samples are ignored in the analysis.

The analysis evaluates the trigger requirement on five time scales with a fixed set of thresholds for the entire 6-year data set. The thresholds were chosen just above the highest hardware levels used during the majority of observations: the thresholds on the trigger time scales are 18 dc (60 ns), 53 dc (180 ns), 119 dc (540 ns), 325 dc (1620 ns), and 874 dc (4860 ns).

Details of the image and timing parameterization are presented next. The parameterization proceeds along the following steps:
 
1) The duration of the pulse is measured by the 1/$e$-width of the summed FADC channels.

\begin{figure}[!ht]
    \includegraphics[width=0.3 \textwidth,clip]{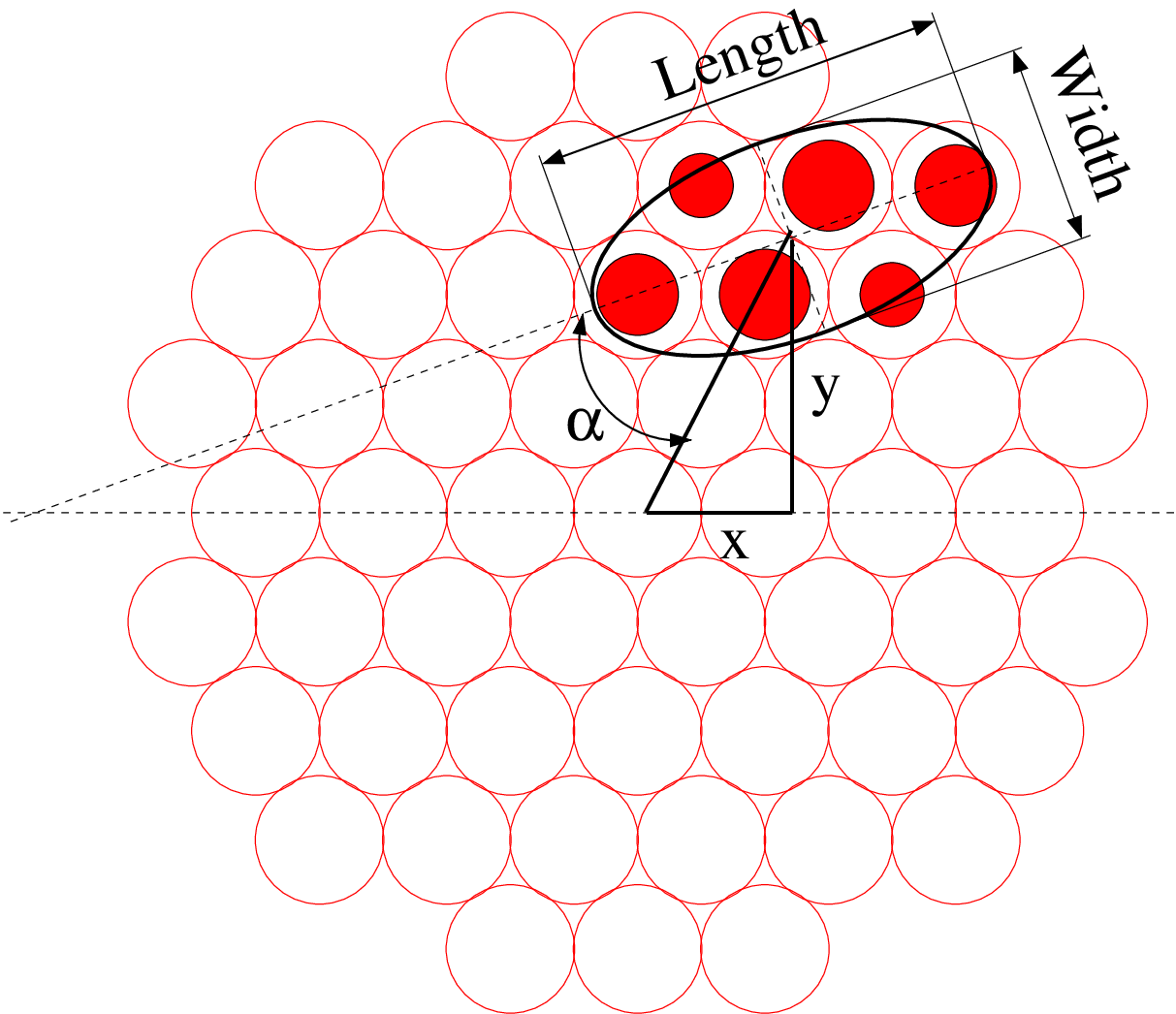}
\caption{Illustration of Cherenkov radiation imaged by SGARFACE. Pixels are shown as thin outlines and the amount of light is represented by filled circles. The shower image is parameterized by a two-dimensional Gaussian distribution with a measured width, length, centroid, and orientation angle, $\alpha$, as indicated. }
\label{fig:hillas}
\end{figure}

2) The event is integrated for the duration of the pulse. The integrated image is characterized by the centroid of the light distribution, the width and length, and the rotation angle, $\alpha$, of the major axis away from the camera center, see Fig.\ref{fig:hillas}. These parameters are calculated from the moments of the light distribution \cite{Hillas1985,Survey1988-91_Reynolds93}. In addition, the peak values of the seven brightest pixels,  $max_{1..7}$, are recorded. The contrast ratio is measured from the ratio of the mean brightness of pixels located at distances less than 0.4\degr\ from the centroid, versus those at distances greater than 1\degr.

3)  The speed of the shower core along the image major and minor axes is measured from the instantaneous centroids of the ten brightest pixels.

\subsection{Monte-Carlo Simulations \label{sec:simulations}}
The sensitivity of SGARFACE is determined from Monte-Carlo simulations. We use the KASCADE~\cite{kascadeKertzman94} air-shower simulation program, version 7.3, together with GrISU(tah) \cite{grisu} to simulate the air-Cherenkov production and telescope response. The calibration of the light-throughput of the telescope and the gain of the PMTs is described in Section \ref{sec:dc_pe}.

\begin{figure}[!ht]
    \includegraphics[width=0.45 \textwidth,clip]{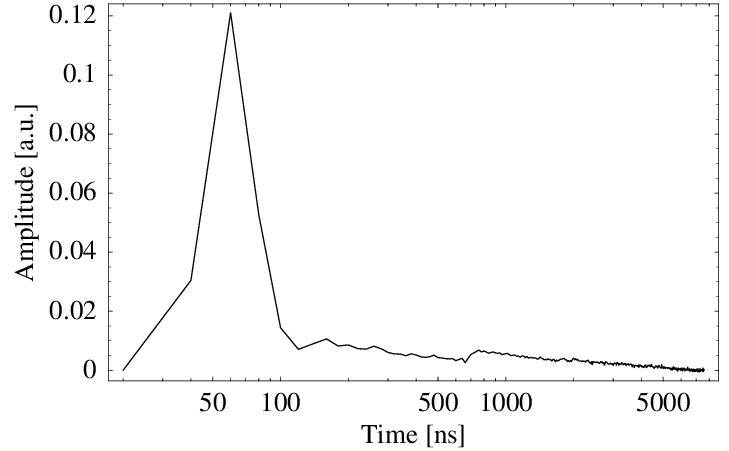}
  \caption{The measured pulse shape of single photo electrons at the input of the FADCs. }
  \label{fig:pulse_shape}
\end{figure}

The simulation of the detector electronics includes the single photo-electron pulse shape received by SGARFACE and is shown in Fig.~\ref{fig:pulse_shape}. The long tail in the pulse shape is caused by the implementation of the AC coupling of the signal cables and their back-termination at the camera. Due to this signal shape, more charge is accumulated on longer integration time scales. This reduces the effective trigger threshold and produces an increase in sensitivity nearly proportional to $\sqrt{\mathrm{time}}$. This was unforeseen in the design of the trigger and helps the performance of the MTSD. The trigger threshold at the shortest time scales corresponds to about 12 Cherenkov photons per m$^2$ per pixel per 60 ns.
\subsection{Background Rejection \label{sec:cuts}}
Data from 2003/4 was used to understand the background events and, in conjunction with simulated bursts of $\gamma$ rays, to develop a background-rejection method. Several sources of background are seen in the data and can be separated into atmospheric (cosmic rays, lightning) and man-made phenomena. 

The largest number of background events are cosmic rays where the Cherenkov radiation is seen at various off-axis directions resulting in event durations of up to 100~ns. SGARFACE also detects events lasting up to $\sim$450~ns where a central bright spot moves through the field of view. These events are compatible with fluorescence emission from cosmic rays.

From comments by the 10~m operators, we determined that lightning produces another type of background seen occasionally in the data. These events have a patchy structure and are longer than 16~$\mu$s, compatible with the measurements by \cite{Lightning1998} that show a duration between 25~$\mu$s and 2~ms. Another type of background that we considered is light from meteor trails. Though the light in the trails has a duration long enough to trigger the MTS, $\sim$0.3~s \cite{Brosch2004_meteors}, the lateral angular extend of $<<$0.1\degr\ ~\cite{Kaiser2004_meteors} is not large enough to trigger the pattern-sensitive unit with a 7-fold coincidence requirement.

The 10~m observing log contains the rating of weather conditions by the observers. We did not use these ratings in cutting run-time a-priori for two reasons: One, the ratings are subjective and two, the GPS time information was not recorded properly in the 10~m run-header files beginning in 2007. The later would have made a cross-correlation of the weather information with SGARFACE events a very labor intensive project. Instead, we decided to use all available data and only investigate the observing conditions for events that passed our selection criteria. The majority of 10~m observations are taken during good or very good night-sky conditions, though occasionally the telescope was operated for short periods of time when the weather was bad. By including all available data, the integrated lifetime of SGARFACE may be too large large by as much as 10\%. We do not apply a compensating lifetime reduction in the analysis.

Background that was identified as man-made includes the light produced by airport/airplane beacons and is detected either directly or from the reflection by haze or clouds in the night sky. Directly visible from the Whipple 10~m telescope is the beacon light at the Nogales airport. We have identified some events as arising from these beacons based on the repetition rate. A third source of background is RF noise picked up by the data acquisition system. This background typically affects only a small fraction of the signal channels, but can cause triggers on long time scales. To eliminate these known sources of background, events are required to pass the ''preliminary'' cuts shown in Tab.~\ref{tab:prelim_cuts}.
\begin{table}
\begin{tabular}{l|r}
Parameter & Cut\\
\hline
\hline
Elevation & $>$20\degr \\
Trigger   & 7-fold \\
max$_1$   & $<$230 dc \\
Pulse stop     & $<$1750 - \emph{W}\\
\end{tabular}
\caption{ Preliminary cuts applied to the entire data set. Here, \emph{W} refers to the width (in 20~ns samples) of the top-hat filter function. }
\label{tab:prelim_cuts}
\end{table}

Following the application of these preliminary cuts to the entire data set, 100,221 events remained. A more selective set of cuts was then developed to reject as much background as possible while retaining all simulated bursts. Bursts were simulated at discrete energies between 100~MeV and 10~GeV, off-axis arrival direction of up to 4\degr, and fluence from 0.175 to 32~$\gamma$-rays/m$^2$. The simulated telescope pointing directions are as shown in Fig.~\ref{fig:el_az}.

\begin{figure}[!ht]
    \includegraphics[width=0.45 \textwidth,clip]{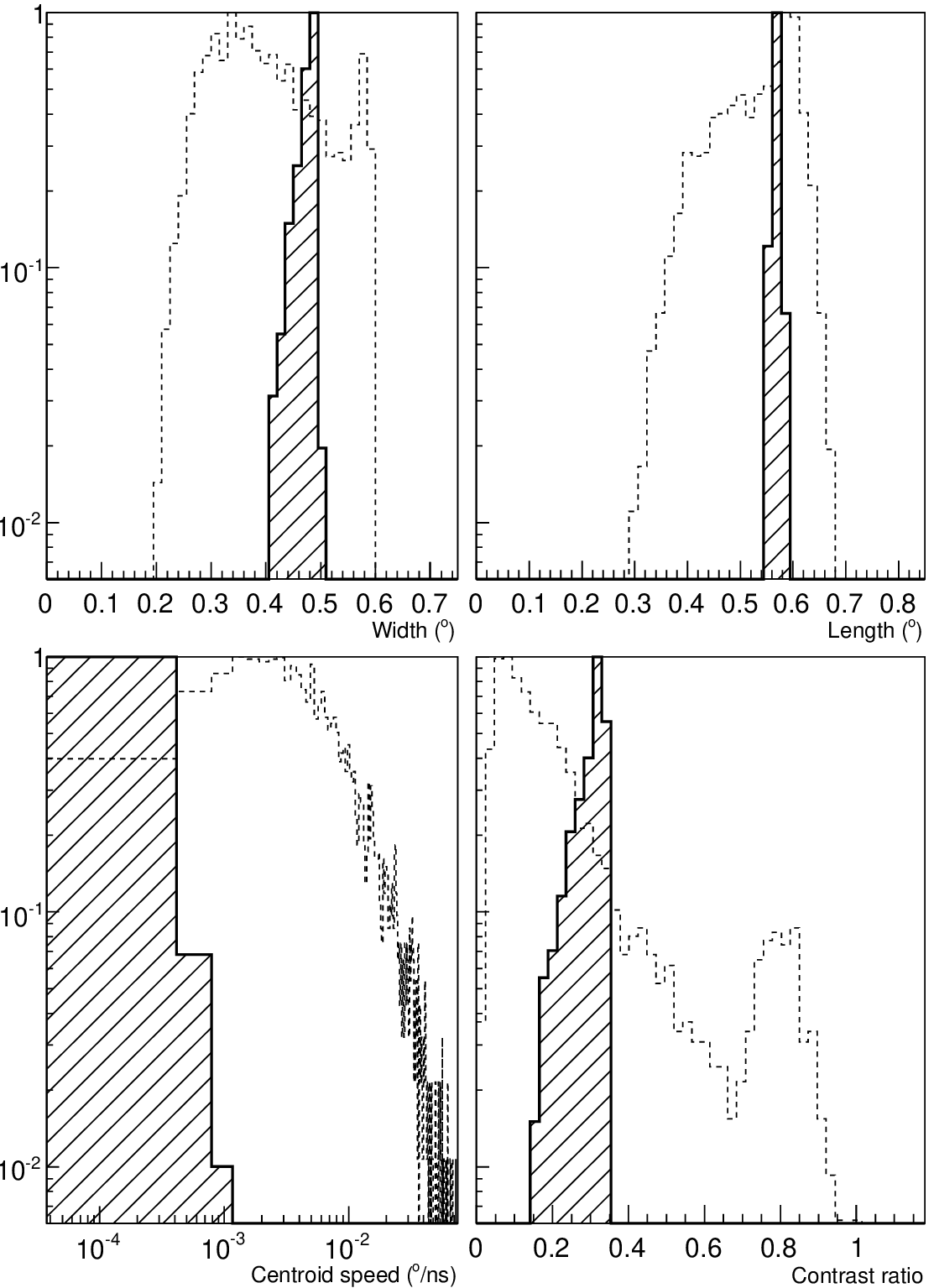}
\caption{Normalized parameter distributions of data taken during 2003 (\emph{dashed}) after application of preliminary cuts and at elevations above 80\degr. The \emph{hatched areas} show the parameter range of simulated bursts of 1~GeV $\gamma$ rays arriving on-axis at zenith with a burst fluence of 2 $\gamma$/m$^2$. Image parameters were calculated using a one-sample wide top-hat filter. The large excess of events in the histogram of contrast ratios $\sim$0.8 is due to flashes of light from a Nitrogen pulser used in the nightly calibration procedure of the 10~m telescope.}
\label{fig:zenith_data_sim_comp}
\end{figure}

For any particular telescope pointing and burst property, the measured burst parameters occupy a narrow space compared to the data. Fig.~\ref{fig:zenith_data_sim_comp} compares the parameter distributions of data taken in 2003 with simulated bursts of $\gamma$ rays for one set of simulation inputs described in the caption. The parameter distributions do not change significantly with the width of the top-hat filter or with the burst duration. In principle, this separation in parameter space between bursts and background could be used to achieve a very efficient background rejection.

\begin{figure}[!ht]
    \includegraphics[width=0.45 \textwidth,clip]{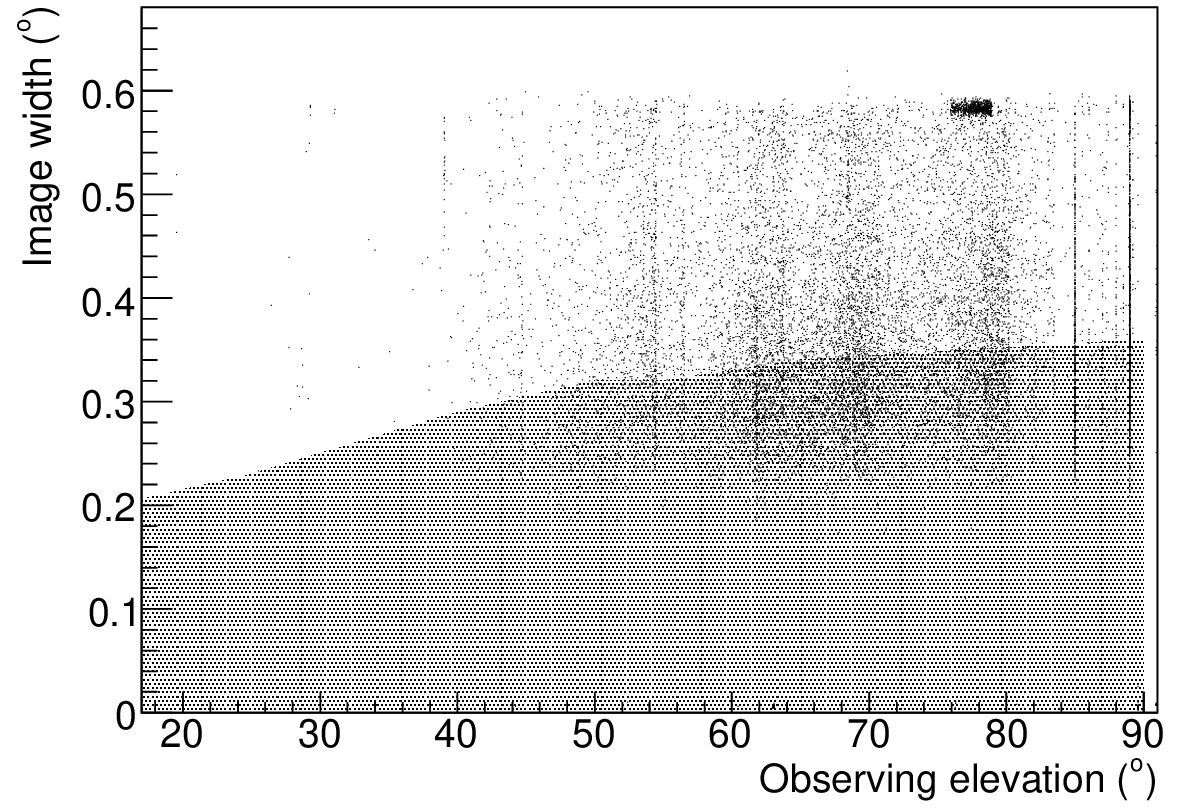}
\caption{ Measured image \emph{width} vs. telescope elevation for events seen in the 2003 data. The \emph{shaded} region shows the rejected parameter space. The excess of events in the upper part of the figure (width of $\sim$0.58\degr\, 78\degr\ observing elevation) are not of astrophysical origin as they occurred while engineering work was performed on one of high-voltage power supplies.  }
\label{fig:width_cut}
\end{figure}

The fluence and location of the burst in the field of view and, at lower elevations the azimuth dependence, create a five-dimensional parameter space for which specific cuts would need to be developed. We do not consider it practical to do so at this time. An alternative, which we also do not implement because of its dramatic effect on the phase space to detect PBH explosions, would be to restrict the data set to a very limited field of view and telescope pointing. Instead, we developed a very simple set of cuts, given in Table~\ref{tab:cuts}, that does not reject any simulated bursts of gamma rays. We note that with these simple cuts we expect some background events to pass the cuts and these events will be investigated a-posteriori. The cut on \emph{width} with telescope elevation is shown in Fig.~\ref{fig:width_cut} and compared to the \emph{width} seen for data. This set of cuts accepts all simulated bursts of $\gamma$ rays and retains 36,457 events out of the entire data set.

\begin{figure}[!ht]
    \includegraphics[width=0.45 \textwidth,clip]{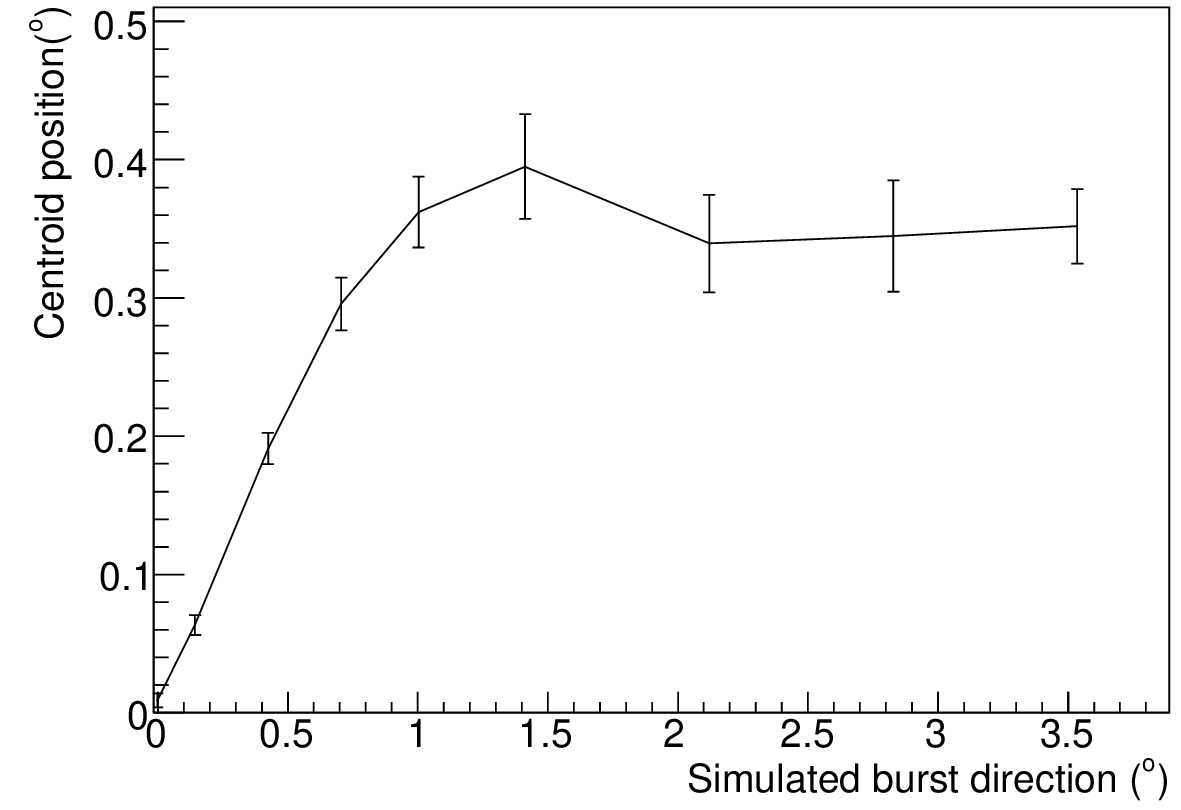}
\caption{ Variation of the measured \emph{distance} of the image centroid from the center of the field relative to the true burst arrival direction. The error bars show the RMS spread of the distribution. }
\label{fig:dist_offset}
\end{figure}

The variation of the mean centroid position, \emph{distance}, with off-axis arrival direction, is shown in Fig.~\ref{fig:dist_offset}. The burst direction can be determined from the \emph{distance} parameter, if the \emph{distance} is less than 0.25\degr. At larger pointing offsets, the field of view limits the positional reconstruction ability. 

\begin{table}
\begin{tabular}{r|r}
Parameter & Cut\\
\hline
\hline
Length & $<$0.65\degr \\
Distance & 0.47\degr ... 0.58\degr \\
Contrast & $<$0.67 \\
\hline
Centroid speed & \\
time scale 1 & $<$0.0276\degr/ns\\
time scale 3 & $<$0.0109\degr/ns\\ 
\end{tabular}
\caption{ Cuts on the parameters that retain all simulated bursts. }
\label{tab:cuts}
\end{table}
\subsection{Detector Sensitivity\label{sec:sensitivity}}
\begin{figure}[!ht]
    \includegraphics[width=0.45 \textwidth,clip]{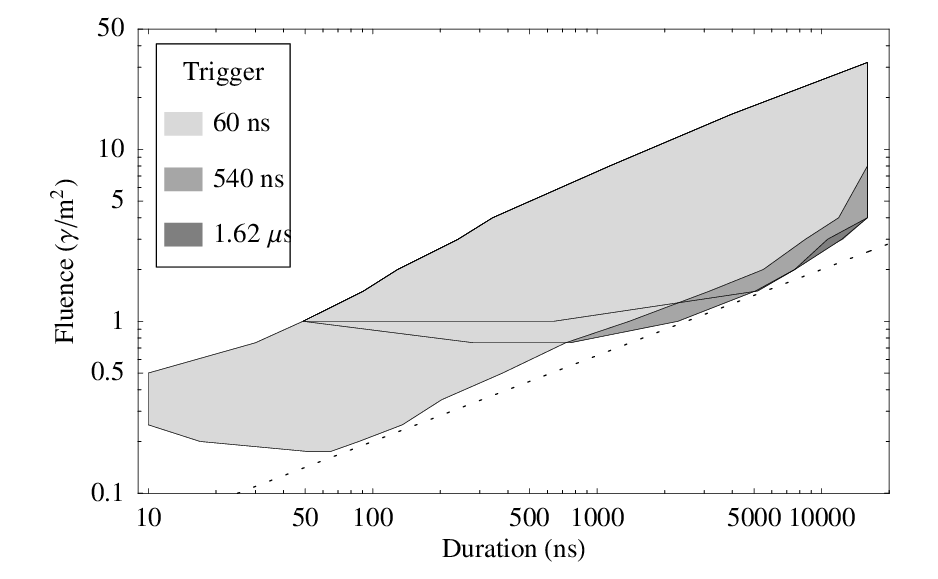}
\caption{ Accepted fluence versus burst duration at various trigger time scales for a 1~GeV burst of $\gamma$ rays observed at zenith and centered in the field of view. Only the three trigger time-scales that increase the acceptance are shown. The solid lines outline the trigger regions corresponding to each timescale. The contribution of the 1.62~$\mu$s trigger timescale is shown by the small, dark shaded area at the far right. The \emph{dotted} line shows the fluence that could be achieved if the trigger were at the noise level, i.e. increasing proportional to $\sqrt{\mathrm{time}}$. }
\label{fig:fluence_sensitivity}
\end{figure}

The fluence sensitivity is determined as a function of burst duration, $\tau$, primary $\gamma$-ray energy, E$_0$, the telescope elevation, $El$, azimuth, $AZ$, and the burst position within the field of view, $\vec{\theta}$. We make the assumption that the sensitivity can be expressed in terms of four independent functions
\begin{equation}
F_{min}[El, AZ, \tau, E_{0}, \vec{\theta}] 
= \frac{F_{min}[\tau]}{g[E_0]\times h[El, Az]\times k[\vec{\theta}] },
\end{equation}
where $F_{min}[\tau]$ is determined after preliminary cuts and at nominal conditions: telescope pointing at zenith, $E_{0}$ = 1 GeV, and the burst is centered in the field of view, $\vec{\theta}$=0. The upper limit on the detectable fluence, $F_{max}$, is due to the limited dynamic range of the FADCs, and it is determined from simulations. The lower and upper limits on the fluence are determined by requiring a 90\% trigger probability. The contribution of each trigger time scale to the fluence sensitivity is shown in Fig.~\ref{fig:fluence_sensitivity}, where only trigger time scales are shown that enhance the sensitivity at longer burst durations. Already at the shortest, 60 ns, time scale, the fluence threshold increases nearly at the noise limit, proportional to $\sqrt{\mathrm{time}}$, as can be seen by comparison with the thin dotted line in Fig.~\ref{fig:fluence_sensitivity}. The systematic uncertainty in the fluence is due to the energy calibration of the detector and is estimated to be 15\%.

To determine the functions $g[E_0]$, $h[El, Az]$, and $k[\vec{\theta}]$, we note that the fluence threshold is proportional to the value of the seventh brightest pixel, max$_{7}$. We use the relative change of max$_{7}$ with respect to $E_{0}$=1 GeV, El = 90\degr, Az=0\degr, and $\vec{\theta}$=0, to determine these functions. The variation with primary energy was already shown in Fig.~\ref{fig:sensitivity_energy}; here it is formally identified with $g[E_0]/E_0$.

\begin{figure}[!ht]
    \includegraphics[width=0.45 \textwidth,clip]{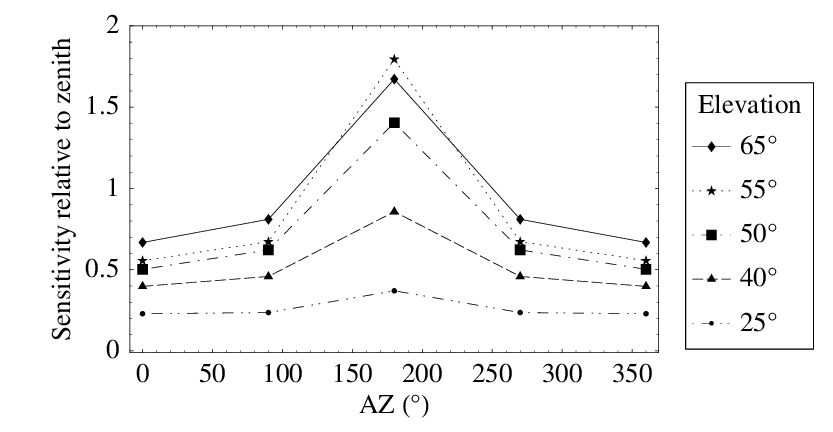}
\caption{ Trigger sensitivity relative to zenith as a function of telescope azimuth and elevation. North is at 0\degr~, East is at 90\degr~AZ. The sensitivity to detect bursts of gamma rays is greatest along the direction of the Earth's magnetic field. At the location of the Whipple 10~m telescope, the Earth's magnetic field points at 170.2\degr\ azimuth and 59\degr\ elevation. Data was provided by the Geomagnetism Department at the National Geophysical Data Center. }
\label{fig:el_az}
\end{figure}

\begin{figure}[!ht]
    \includegraphics[width=0.45 \textwidth,clip]{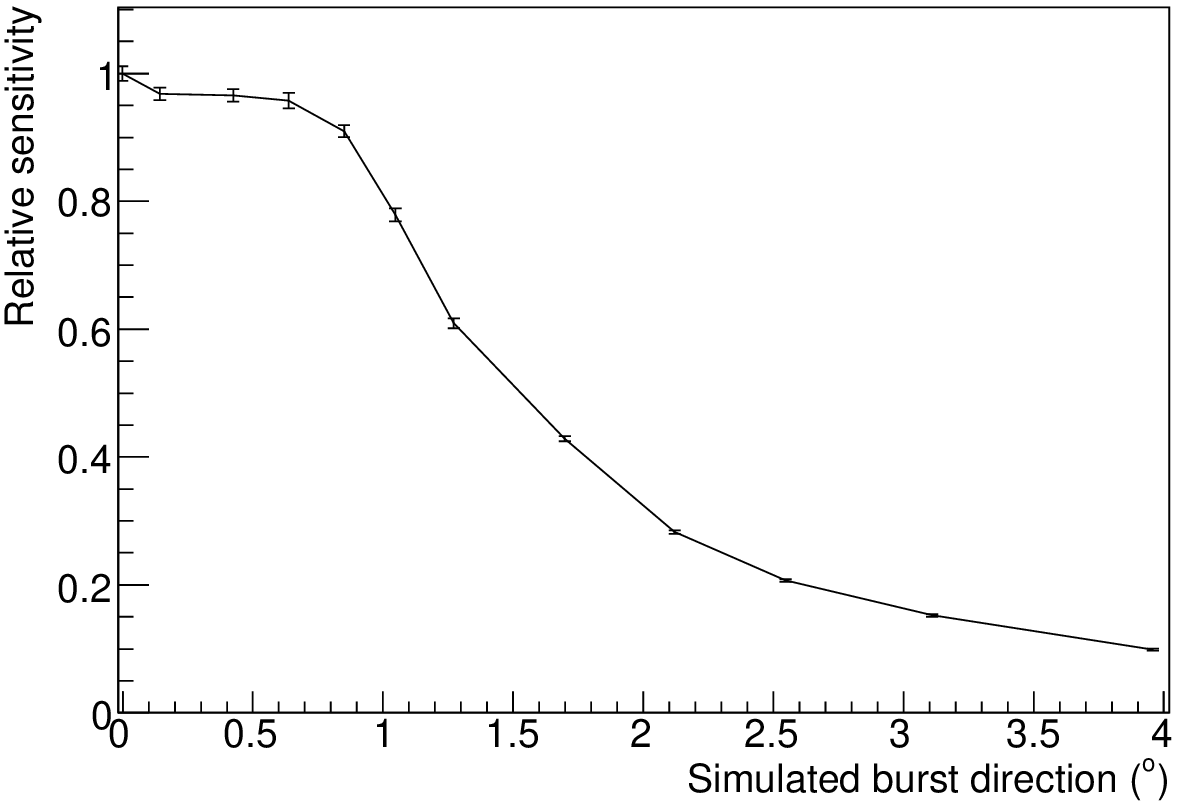}
\caption{ Variation of the sensitivity to detect bursts arriving off-axis. The arrival direction is along a line inclined by 45\degr\ with respect to the x-axis. In the analysis, we limit the half-field of view to $<$3\degr. }
\label{fig:dist_max7}
\end{figure}

Cherenkov images from bursts of $\gamma$ rays at 1 GeV are of 2 to 3\degr\ angular extend, as shown in Fig.~\ref{fig:Cherenkov_distribution}. SGARFACE has a geometric field of view of 2.4\degr\ and is sensitive to bursts originating well outside of this range. As burst images are slightly oblong due to the Earth's magnetic field, the sensitivity decreases faster for an offset parallel the direction of the Earth's magnetic field than perpendicular to it, see~\cite{Sgarface_Lebohec2005}. We ignore this directional dependence and use the average value to characterize the off-axis sensitivity; i.e. $k[\vec{\theta}]$ = $k[\theta]$, only. The relative sensitivity with respect to the burst direction in the field of view is shown in Fig.~\ref{fig:dist_max7}. The sensitivity variation with telescope pointing, $h[El, AZ]$, shown in Fig.~\ref{fig:el_az}, is due to shower geometry, Cherenkov-light absorption in the atmosphere, and the effect of the geomagnetic field.

The functions, $g$, $h$, and $k$ are also used in scaling the upper limit, $F_{max}$. This is not strictly correct as the maximum fluence is limited by the brightest pixel, max$_{1}$, and not max$_{7}$. However, the introduced error is estimated to be small, 5-10\%, and the impact is negligible on the total volume for detection of PBH explosions.
\subsection{Detection Volume for PBH Explosions\label{sec:volume}}
\begin{figure}[!ht]
    \includegraphics[width=0.45 \textwidth,clip]{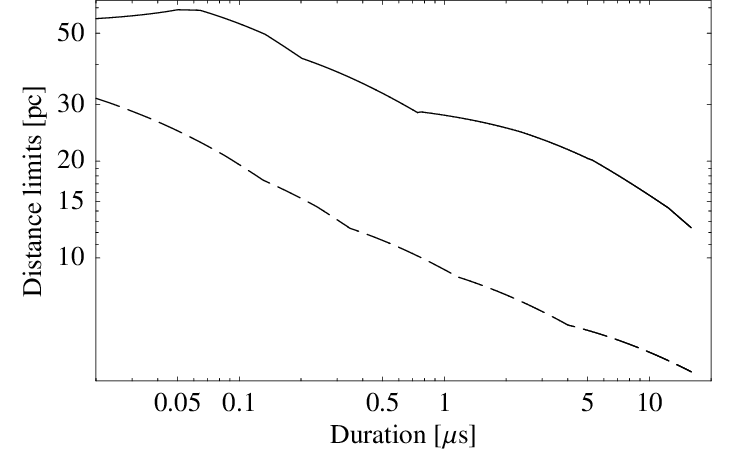}
\caption{Maximum, $R_{max}$, (\emph{solid}) and minimum, $R_{min}$, (\emph{dashed}) detectable distance for PBH explosions observed at the center of the field of view and telescope pointing at zenith. The total burst energy is $7.5\times 10^{45}$ eV released in the form of 1 GeV $\gamma$ rays. }
\label{fig:distance}
\end{figure}

\begin{figure}[!ht]
    \includegraphics[width=0.45 \textwidth,clip]{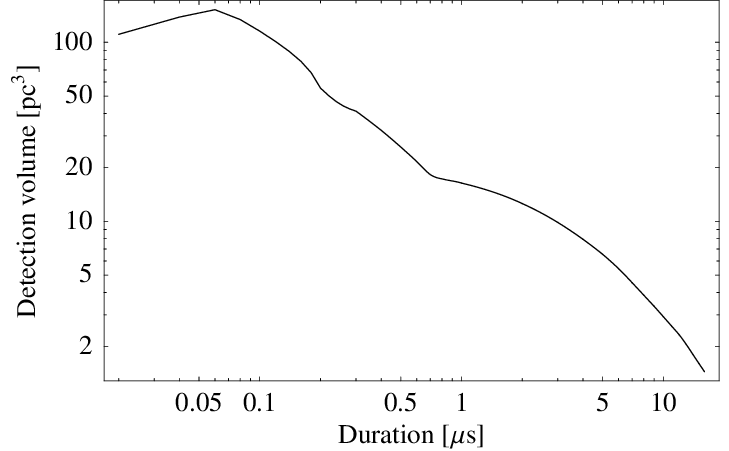}
\caption{Volume for detection of PBH explosions with telescope pointing at zenith. The total burst energy is $7.5\times 10^{45}$ eV released in the form of 1 GeV $\gamma$ rays. }
\label{fig:volume}
\end{figure}

The volume to detect PBH explosions is $\mathrm{d}V=(R_{max}^3-R_{min}^3) \mathrm{d\Omega}/3$, where $\mathrm{d}\Omega = \mathrm{d}\phi \sin[\theta]\mathrm{d}\theta$ is the solid angle, and $R_{min,max}$ are the detection limits on the distance compatible with $F_{min,max}$. The distance limits to detect bursts at nominal conditions are shown in Fig.~\ref{fig:distance}. The fluence, $F$, of a burst at distance, $R$, and initial particle number, $N_{0}$ is $F=N_{0}/(4\pi R^2)$, hence the volume scales with $F$ as $V=\mathrm{d}\Omega/(3(4\pi)^{3/2}) \times (N_0/F)^{3/2}$. The initial number of particles can be written in terms of the total energy released in $\gamma$ rays, $E_{T}$: $N_0=E_T/E_0$. The detection volume for PBH explosions can then be written as
\begin{multline}
\mathrm{d}V=\frac{\mathrm{d}\Omega}{3(4\pi)^{3/2}}( \frac{E_T\times g[E_0]\times h[El, Az]\times k[\theta]}{E_0}  )^{3/2} \\
\times ( F_{min}[\tau]^{-3/2} - F_{max}[\tau]^{-3/2} ).
\label{eq:volume}
\end{multline}
The total volume to detect bursts, integrated over a 6\degr\ field of view, is shown in Fig.~\ref{fig:volume} as a function of burst duration. The size of the field of view was chosen somewhat conservatively so that the sensitivity at the outer edge is at least 20\% compared to the center. The systematic error of the volume is 23\%, stemming from the uncertainty in the energy calibration of the detector.
%%
%
%
%
%%%%%%%%%%%%%%%%%%%%%%%%%%%%%%%%%%%%%%%%%%%%%%%%%%%%%%%%%%%%%%%
%
%
\section{Search for PBH Explosions \label{sec:bursts}}
Most events trigger on several time scales, as shown by the overlap of trigger regions in Fig.~\ref{fig:fluence_sensitivity}. In searching for PBH explosions, events are identified using only the parameters of the shortest triggered time scale. An exception is made for very short events of duration less than 60 ns, which only trigger on longer time scales due to the tail of the single pe-distribution. These short events are removed a posteriori by a special cut consisting of \emph{max$_7$}$>$ 100~dc and duration $<$60~ns.

\begin{figure}[!ht]
    \includegraphics[width=0.4 \textwidth,clip]{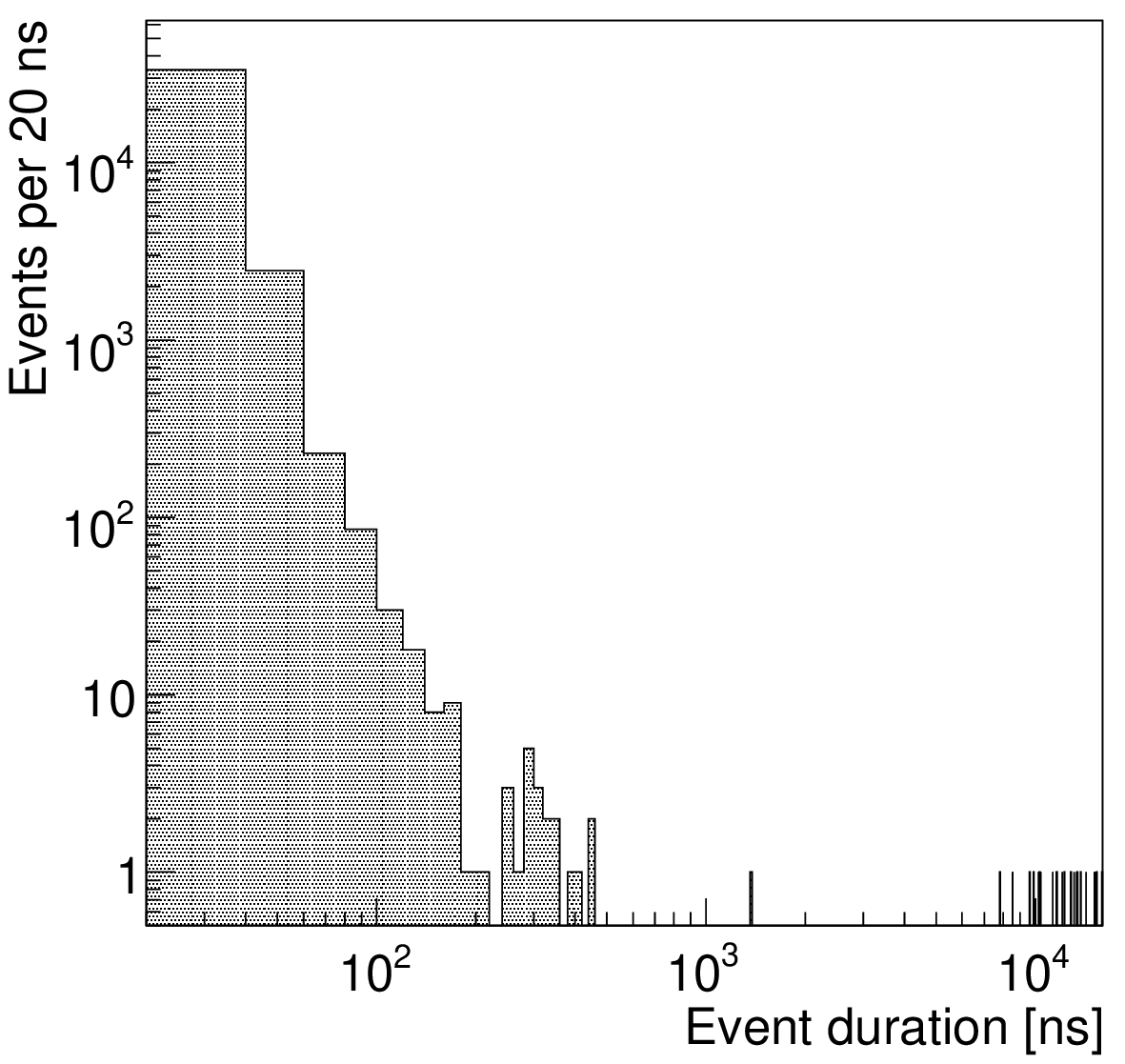}
\caption{ Histogram of the duration of the 36,170 events passing all cuts described in the text. }
\label{fig:duration_ultraShort}
\end{figure}

The duration of events passing all cuts is shown in Fig.~\ref{fig:duration_ultraShort}. Due to the simple parameters and loose cuts, a number of background events pass the cuts. Instead of further refining the cuts at this point, we describe the most likely origin of the remaining events. The hypothesis that the events could \emph{only} have originated from a burst of gamma rays is tested by investigating the image and time-development in more detail and considering the circumstances under which the event occurred: weather, tracking/slewing, comments by observer in the nightly log.

Three distinct populations are visible in Fig.~\ref{fig:duration_ultraShort}. The largest number of events are very short and bright cosmic rays of at least 5 TeV and $>$3\degr\ arrival direction with respect to the telescope pointing. A second detector separated by at least a few hundred meter would be needed distinguish cosmic rays from bursts of $\gamma$ rays. A second population peaks near 300~ns, while a third group of events is seen to have durations longer 7~$\mu$s.

\begin{figure}[!ht]
  \includegraphics[width=0.2 \textwidth,clip,angle=-90]{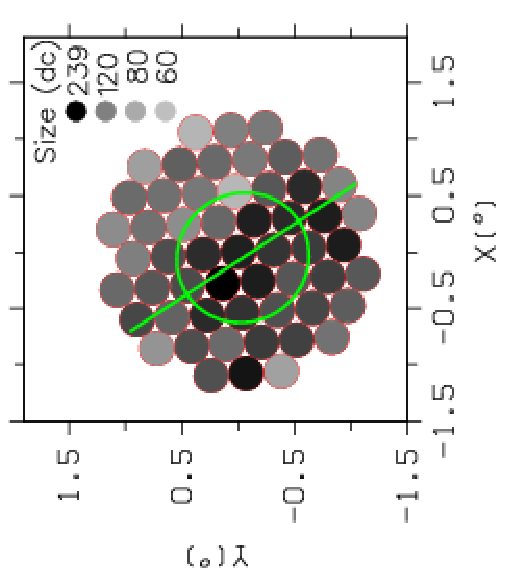} 
  \includegraphics[width=0.2 \textwidth,clip,angle=-90]{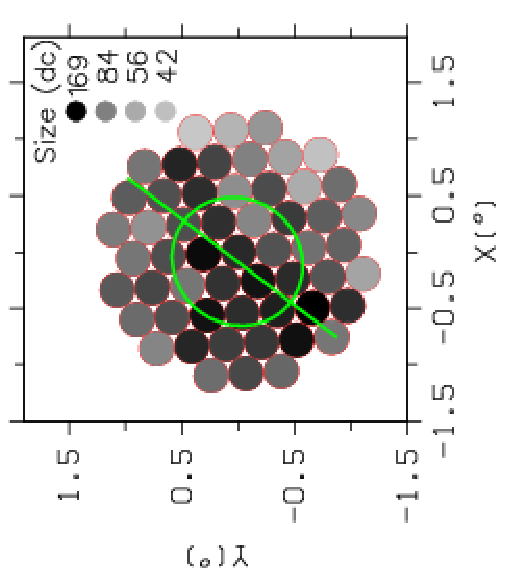} 
\caption{ Two unusual events that exhibit a very broad light distribution and appear to be centered in the field of view. The time development of the events was uniform and they lasted approximately 140 ns. The fitted ellipsoid and the direction of the major axis are superimposed on the camera pixels. }
\label{fig:broadEvents}
\end{figure}

\begin{figure}[!ht]
    \includegraphics[width=0.4 \textwidth,clip]{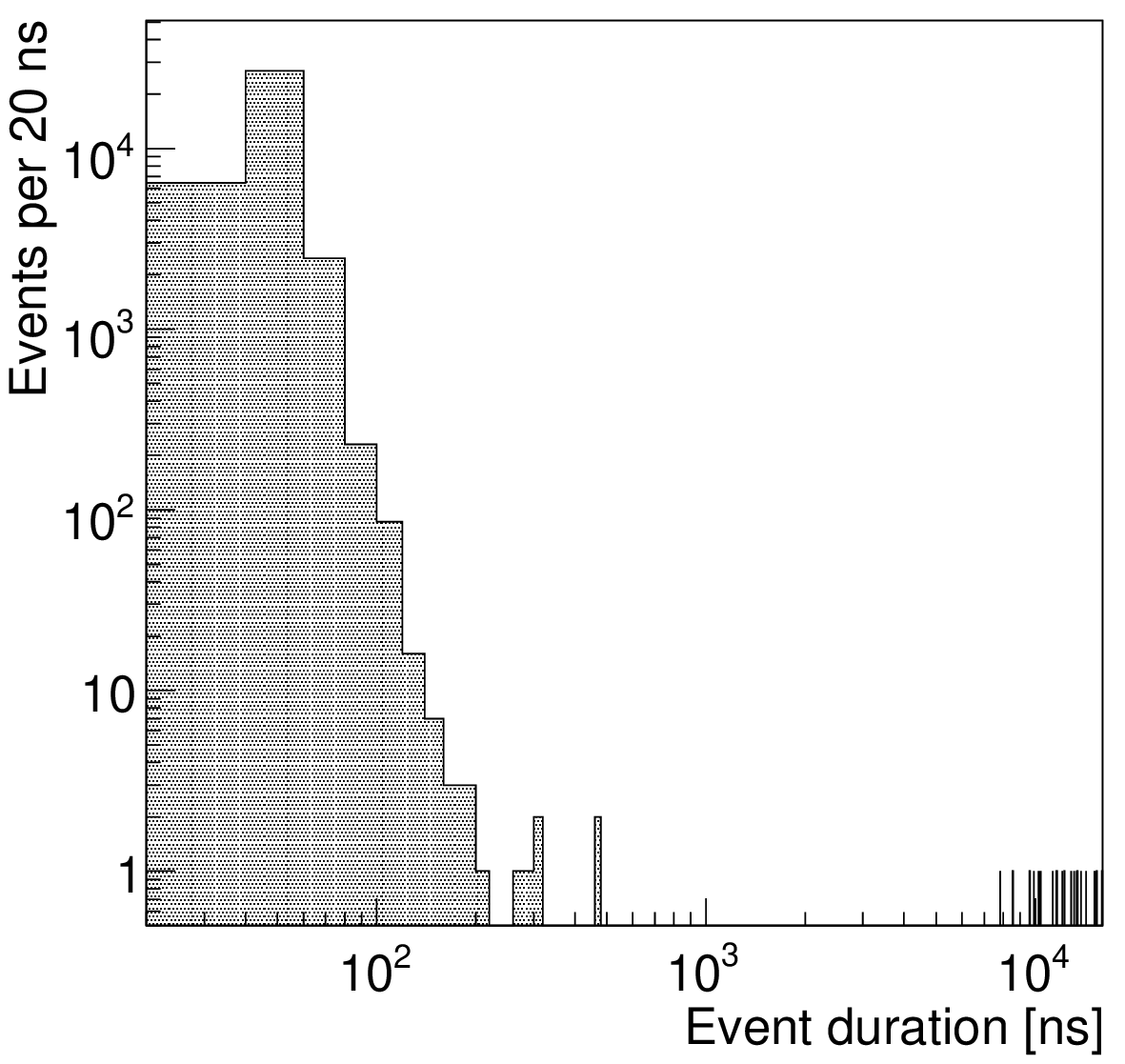}
 \caption{ Histogram of event durations after removal of events visually identified as having a compact spot moving through the field of view. It is likely that these removed events were produced by fluorescence emission of cosmic rays. }
\label{fig:duration}
\end{figure}

Next, details are given about the second population of events, those of duration between 120~ns and 500~ns in Fig.~\ref{fig:duration_ultraShort}. The time-resolved analysis of the 87 events contained in this group reveals that a large fraction exhibit a characteristic spot of $\sim$0.3\degr\ diameter moving through the field of view at $\sim5\times 10^{-3}$~\degr/ns. The three events that were detected at less than 45\degr\ elevation develop in a downwards direction in the field of view. Events detected near zenith appear to have random directions. If the detected radiation of these events is fluorescence emission from a cosmic ray traveling perpendicular to the line of sight, then the total light collected by SGARFACE corresponds to primary energies between $\sim2.5-25\times10^{17}$~eV~\cite[pg. 288-290]{Longair1992}. Because the timing and image parameters were calculated using a moment analysis, it is difficult to distinguish the motion of a compact spot from the image of a burst of $\gamma$ rays seen largely off-axis. A more sophisticated analysis coupled with cuts on the timing gradient as a function of off-axis distance and image brightness would likely have eliminated these events. After identifying by eye 51 events that clearly show the characteristic spot and removing those from the data set, the event durations of the total data set, corresponding to 36,118 events, are shown in Fig.~\ref{fig:duration}. Of the remaining events in the second population, 22 show a light distribution that peaks at the camera edge and slowly moves in time, making a burst origin unlikely. The other 14 events satisfy the criteria for burst images; that is they exhibit a broad image shape and uniform time development. Further investigation shows that 12 of these events were seen mainly off-axis, resulting in ambiguity as to the determination of their origin. It is possible that all events have were caused by cosmic-ray fluorescence emission with the detected variations being due to differences in energy and the viewing angle. The two remaining events occurred at 10:47 UTC, 20 April 2007 and at 08:03 UTC, 23 April 2007. The events show an extremely broad angular light distribution, $>2$\degr, and uniform time development with rise time of 20~ns, fall time of 40~ns, and total duration of $\sim$140~ns.  The time-integrated images are shown in Fig.~\ref{fig:broadEvents}. The first event occurred during routine observations of Mrk~501 at 80.5\degr~El and 332\degr~AZ, and the second event was recorded just before the nightly calibration (zenith) run was carried out with the telescope pointed at 80\degr~El and 0\degr~AZ; corresponding to a random direction in the sky. They occurred during very good night sky conditions and no detector malfunction could be identified. SWIFT detected a gamma-ray burst at 6:18 UTC, 20 April, but no burst was detected on 23 April. We note that atmospheric fluorescence from a burst of X-rays could produce short events with a very broad angular distribution~\cite{Elliot1972}. However, we do not claim to have detected bursts of $\gamma$ rays or of X-rays, because we cannot completely rule out atmospheric phenomena.

\begin{figure}[!ht]
    \includegraphics[width=0.48 \textwidth,clip]{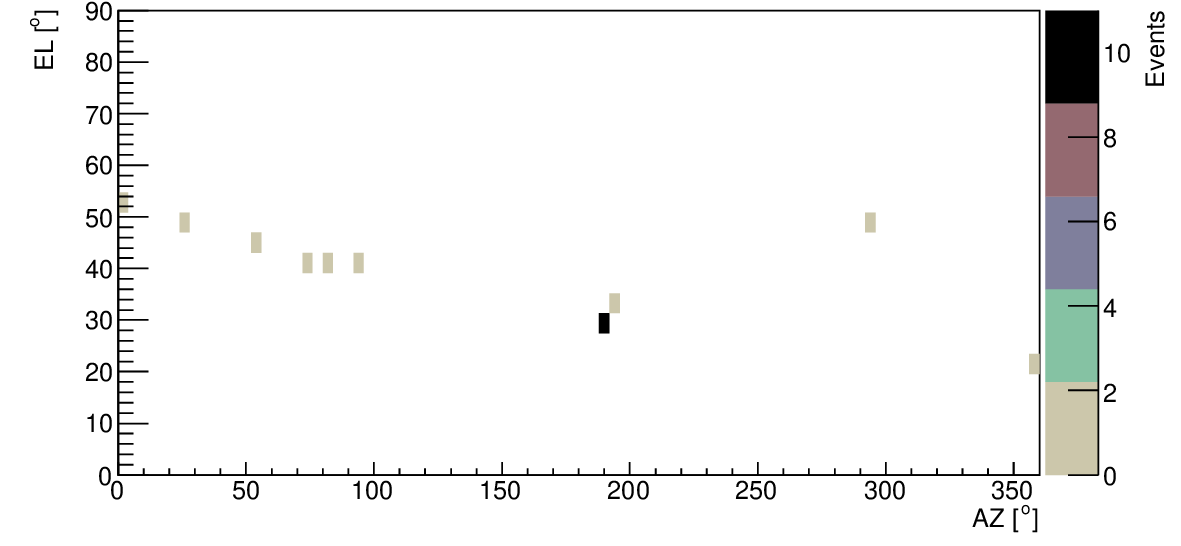}
\caption{ Map showing the origin of the 24 events detected with durations longer than 7~$\mu$s. }
\label{fig:longEventsMap}
\end{figure}

Durations longer than 7~$\mu$s were measured for 24 events. The arrival direction of the events is shown in Fig.~\ref{fig:longEventsMap}. Twelve of these events occurred within 3 minutes of each other on 10 November, 2007, with the telescope pointing South at 28\degr\ elevation during very bad night-sky conditions: distant lightning was noted in the observer log. Unfortunately, it was not possible to filter out such data automatically because, as mentioned in Sect.~\ref{sec:cuts}, the GPS time stamps in the 10~m database that are necessary to correlate the observer's weather-rating with individual SGARFACE events were missing since 2007.

\begin{figure}[!ht]
  \includegraphics[width=0.3 \textwidth,clip,angle=-90]{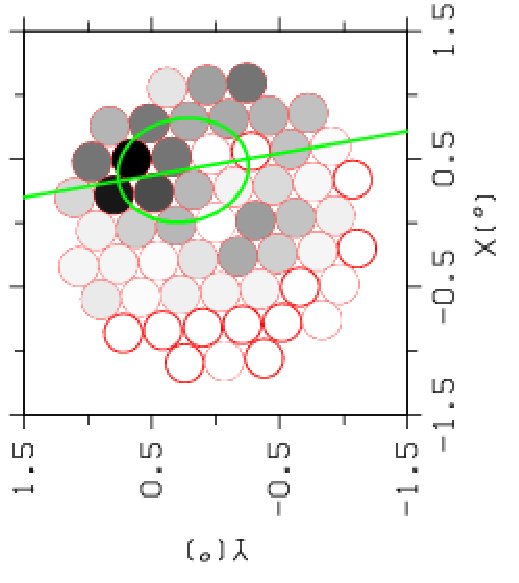} 
\caption{ An event of 9.5~$\mu$s duration that passes all cuts but exhibits patchy structure that is not expected from a burst of $\gamma$ rays. }
\label{fig:patchyEvent}
\end{figure}

During the previous night and under good night-sky conditions, six of the 24 events occurred within one minute of each other, while the telescope was slewing to a new observing position. These events were probably caused by RF noise generated by the telescope tracking system. Another two events occurred during engineering operation while the telescope was not scheduled for astronomical observations. One event occurred during a run flagged as bad by the observers due to a high trigger rate in part of the camera. Another event, Fig.~\ref{fig:patchyEvent}, shows patchy structure and is incompatible with the smooth distribution of the integrated Cherenkov light expected from a burst of $\gamma$ rays, shown in Fig.~\ref{fig:Cherenkov_distribution}. In any case, these events are excluded because the 10~m was not performing astronomical observations. The remaining two events show a rising time-development past the initial drop-off from the peak and are longer than the 16~$\mu$s FADC depth. These events passed the analysis cuts because only the initial peak is parameterized while the subsequent time development is ignored. A possible explanation for these events is distant, high-altitude lightning. In summary, no events detected at the longest durations can be explained solely by a burst of $\gamma$ rays.

\begin{figure}[!ht]
    \includegraphics[width=0.48 \textwidth,clip]{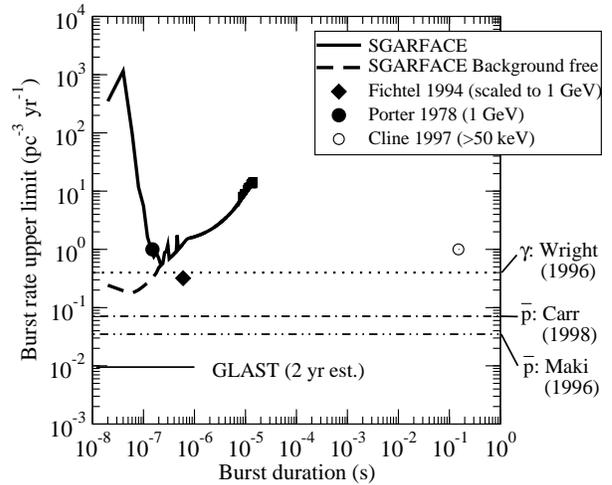}
\caption{99\% upper limits placed on the density of PBH explosions by SGARFACE and by other experiments. All direct-detection limits have been scaled to a total energy of $7.5\times 10^{45}$ eV released in 1 GeV $\gamma$ rays. The background-free SGARFACE limits are shown with a \emph{thick dashed} line. Indirectly determined upper limits from background measurements are not necessarily at the 99\% CL level due to large systematic uncertainties in the galactic halo enhancement. }
\label{fig:limits}
\end{figure}

In conclusion, based upon our study of events identified as cosmic-ray background, atmospheric phenomena, or events triggered by RF noise, we cannot exclude that no bursts of $\gamma$ rays have been detected and that the entire excess is due to background. Upper limits on the true PBH explosion rate, shown in Fig.~\ref{fig:limits}, were calculated using Poisson statistics of the remaining background events shown in Fig.~\ref{fig:duration}, where an exposure time of 1502 hours was used. This exposure time arises from correcting the true lifetime of 2267 hours for sensitivity changes with elevation and optical throughput efficiency. The upper limits can be scaled to different total and primary energies using Eq.~\ref{eq:volume} and Fig.~\ref{fig:sensitivity_energy}.
%
%
%%%%%%%%%%%%%%%%%%%%%%%%%%%%%%%%%%%%%%%%%%%%%%%
%
%
%
\section{Discussion and Summary}
The direct search for PBH explosions with SGARFACE covered a broad range of $\gamma$-ray energies and burst durations compatible with a Hagedorn-type black hole evaporation. The experiment is collecting data since March 2003 and makes use of time-resolved images of Cherenkov radiation to distinguish the candidate bursts of $\gamma$ rays from the much more numerous cosmic-ray initiated showers. In the analysis, the detector was modeled in detail and Monte-Carlo simulations of bursts of $\gamma$ rays were used to determine the astrophysical volume over which PBH explosions could be seen. Using only a single imaging telescope, this is the first time such a detailed search has been carried out and a high background suppression of 98.4\% was achieved. The majority of the remaining cosmic-ray background events are indistinguishable from burst events with a single telescope. Figure~\ref{fig:limits} shows the upper limits that could have been achieved in the absence of the cosmic-ray background, i.e. if a second, identical detector had been placed at a large distance and operated in coincidence. Originally, SGARFACE was designed when the camera on the 10~m camera had a 4.8\degr\ field of view. With this larger field of view an almost three-fold increase in PBH detection volume would have been achieved. 

\begin{figure}[!ht]
    \includegraphics[width=0.45 \textwidth,clip]{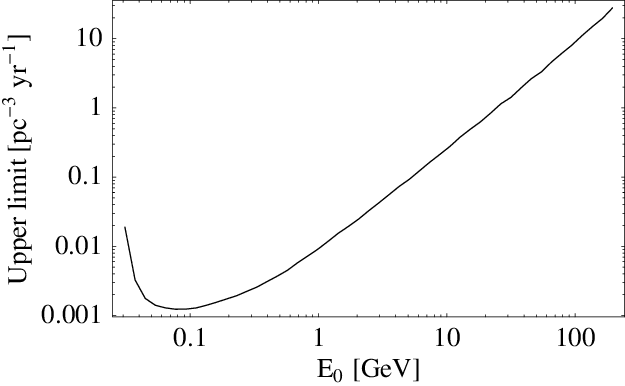}
\caption{ Upper limits on the PBH explosion rate that GLAST can set in 2 years of observation versus energy of the emitted $\gamma$ rays, $E_0$. The total burst energy released in $\gamma$ rays is $7.5\times 10^{45}$ eV. }
\label{fig:glast_limits}
\end{figure}

The GLAST satellite is sensitive to bursts of $\gamma$ rays above 0.03~GeV energy. While the detector has a dead-time of 26~$\mu$s, multiple photons can be detected within the $\sim$ 1~$\mu$s coincidence gate width, enabling a direct search for bursts of $\gamma$ rays. In determining the fluence sensitivity of GLAST to sub-$\mu s$ burst of $\gamma$-rays, we require a minimum of 10 photons to be collected. Assuming that no events are detected, we show the 99\% CL upper limits on the PBH explosion rate for a 2 year exposure in Fig.~\ref{fig:limits}. Shown in Fig.~\ref{fig:glast_limits}, is the 99\% upper limit over the entire energy range of GLAST, where the total burst energy was taken as before as $7.5\times 10^{45}$ eV. GLAST is most sensitive to burst of $\gamma$ rays at $\sim$100 MeV due to the relation between sampled volume and $\gamma$-ray energy at a fixed total burst energy.

In comparing the upper limits shown in Fig.~\ref{fig:limits}, it should be kept in mind that different regions of space are sampled by the different detection techniques. For example, GLAST is limited in distance to 14 pc to detect our nominal burst of $\gamma$ rays, while SGARFACE probes PBH explosions out to 60 pc. Limits from cosmic particle backgrounds measure the PBH distribution on the scale of the Galaxy, while relying on the assumed mass distribution function, galactic halo enhancement, and in some cases the cosmic-ray propagation model, to set limits on the current rate of PBH explosions. These different techniques complement each other in searching for all possible signatures of PBH evaporation.

In this paper we considered bursts of $\gamma$ rays to be produced during the final, explosive evaporation of PBHs. In itself, a burst detected with SGARFACE would not be sufficient to claim the detection of a PBH explosion, as bursts might be produced by the emission of $\gamma$ rays contemporaneous with giant radio pulses~\cite{Lyutikov2008} or perhaps via some other, as yet undiscovered, mechanism. The smoking gun to claim the detection of a Hagedorn-type PBH explosion would include measurements of the neutrino and graviton emission, as well as radio and optical observations~\cite{Rees1977,Heckler1997}, but see \cite{MacGibbon2008} for a refutal on such emission. Future searches for bursts should strive towards such a multi-messenger approach.
\section{Outlook}
SGARFACE is background limited and an unambiguous detection can only be made if the cosmic-ray background can be suppressed. The coincidences from a second telescope separated by at least a few hundred meters can be used to veto events caused by cosmic-ray initiated showers and other local phenomena. The VERITAS array of air-Cherenkov telescopes \cite{Veritas2006} is located 7 km from the Whipple 10~m telescope. While SGARFACE has a lower fluence sensitivity than VERITAS  to detect $\mu$s-bursts, the data from simultaneous observations with the two instruments can be used to reject short cosmic-ray events. The analysis of these simultaneous observations is forthcoming and will focus on a search for bursts of $\gamma$ rays from point sources, such as the Crab pulsar.

The sensitivity of the air-Cherenkov technique to detect PBH explosions can be improved by implementing the imaging, multi-timescale trigger pioneered with SGARFACE, on an array of telescopes. With a telescope separation of $\sim$1~km, a global array trigger system would operate background free and reduce the single-telescope trigger threshold. A larger mirror area would further increase the sensitivity, while a wide field of view increases the volume searched for PBH explosions. For example, if the array consists of telescopes with 20~m diameter, 6\degr\ field of view, and improved noise characteristics, the volume over which PBH explosions can be detected increases by a factor of 20 to 30 over SGARFACE.
%
%%%%%%%%%%%%%%%%%%%%%%%%%%%%%%%%%%%%%%%%%%
%
%
\begin{ack}
We thank the VERITAS Collaboration and the staff at the Whipple Observatory for operating and maintaining the 10~m project. We thank Roy McKay, Harold Skank, and Gary Sleedge at ISU for their help in the electronics design and testing. We would also like to thank Jianwei Qiu at ISU for discussing with us the physics of PBH evaporation and the unknown referee for valuable comments and criticism. FK acknowledges support for the SGARFACE project by the Department of Energy, High Energy Physics Division, through the Outstanding Junior Investigator program and generous financial support by Iowa State University.
\end{ack}
%
%
%%%%%%%%%%%%%%%%%%%%%%%%%%%%%%%%%%%%%%%%%%%%%%%%%%%%%%%%%%%%%%%%%
%
\section{Appendix}
The accuracy to which the energy response of SGARFACE can be calibrated depends on the performance and stability of the optical throughput of the telescope, the photo-electron to digital-counts conversion ratio, and the uniformity of the camera response across the field of view.

\subsection{Calibration of the Light-Throughput \label{sec:dc_pe}}
\begin{figure}[!ht]
    \includegraphics[width=0.21 \textwidth,clip,angle=-90]{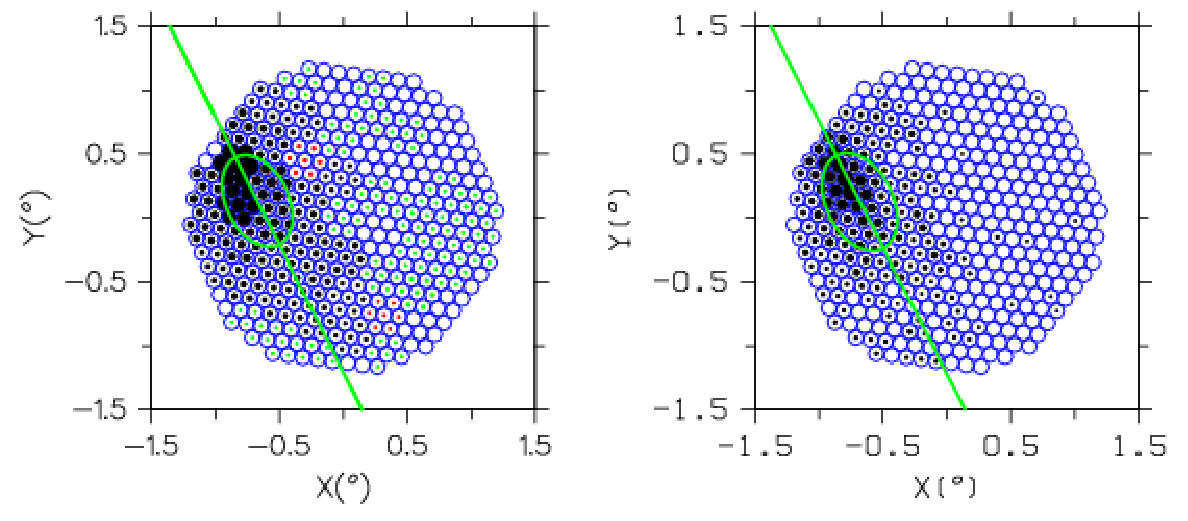}
\caption{A cosmic-ray event detected simultaneously with SGARFACE (\emph{left}) and Whipple (\emph{right}). The total charge is represented by the size of the shaded area in each pixel. The extend of the fitted two-dimensional Gaussian is shown in each image by an ellipse. }
\label{fig:cr_example}
\end{figure}

The telescope response to Cherenkov light was calibrated by direct measurement of all components involved. From previous measurements of the Whipple 10~m electronics and the design of the SGARFACE DAQ, the expected digital-count (dc) per photo-electron (pe) is $0.92\pm0.1$, where the error is due to systematic uncertainties of some 10~m measurements. As these measurements were done only once and at separate times, a second method was developed to track instrumental variations. It consists of two parts: 

1) Events that are recorded simultaneously by both, Whipple and SGARFACE, are used to cross-calibrate the light-level between each cluster of pixels. Cosmic rays produce coincident events about 70\% of the time, providing a large enough statistical sample to perform a nightly calibration. One such coincident event is shown in Fig.~\ref{fig:cr_example}. The ratio of the total light measured in digital counts by Whipple versus SGARFACE is $3.54\pm0.17$, and it is very stable over the five-year time span. The statistical error is due to gain variations between pixels.

\begin{figure}[!ht]
    \includegraphics[width=0.4 \textwidth,clip]{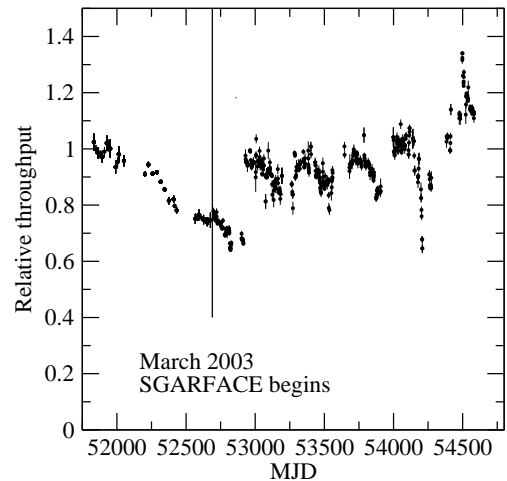}
\caption{Light throughput measured relative to the calibrated point of 0.93 dc/pe in November 2000. }
\label{fig:Sgarface_dcpe}
\end{figure}

2) The light-throughput of the Whipple 10~m is measured from recorded images of complete muon rings. Cherenkov light detected from complete muon rings measures the quality of the local, $<$ 500 m, atmosphere and the telescope optics and electronics throughput. This includes gain changes in the PMT. The periodic variations in the measured light throughput seen in Fig.~\ref{fig:Sgarface_dcpe} is due to aging of the PMTs followed by subsequent increase of the applied high voltage. Because of uncertainties in the ultra-violet response of the mirrors and PMTs, the dc/pe value was tied to the directly measured value of 3.3$\pm$0.3~dc/pe for the 10~m in November 2000 instead of using the absolute value derived from the muon calibration.

Using this cross-calibration method, the dc/pe ratio was determined to be 0.93$\pm$0.1~dc/pe for SGARFACE at the time of the directed measurement and is therefore in agreement with the value obtained from the direct measurements. We note that two channels (18 and 54) exhibit large variations in the calibration ratio and we found this to be due to either, one, or two channels dropping out in the Whipple DAQ.
\subsection{Stability of the FADC Pedestals \label{sec:hardware_stability}}
The SGARFACE FADC input is biased by pull-up resistors to avoid zero-crossing of the digital values. This pedestal level is measured for every channel on an event-by-event basis from the first 750 samples with value below 40 dc.Typical pedestal values are 10-30 dc. Pedestal variations arise from the night-sky background. The pedestal level may also drift with temperature and noise in the data-acquisition system. 

The RMS pedestal fluctuation, as measured by the standard deviation of the distribution during an event, lies between 1.5~dc and0 4.5~dc, depending on night-sky condition and PMT gain. Changes in the mean pedestal value between consecutive events show a spread of 1.3 dc RMS. Over the long term, the pedestals were very stable; all channels showed less than 2~dc drift over the five year analysis period. A minor hardware problem was identified where on rare occasions, traces would start at 0 dc and remain at that level for up to entire stack length. The analysis filters out rare instances of missing FADC traces.
\subsection{Camera Pixel Gain \label{sec:gain}}
The sensitivity across the field of view of the Whipple 10~m camera is set by the gain of individual PMTs. To achieve a uniform gain, the high-voltages of the PMTs are adjusted at the beginning of each observing season. In addition, the Whipple high-energy $\gamma$-ray observations are flat-fielded in software with an auxiliary calibration run taken each night. For this purpose, a Nitrogen flasher is used to uniformly illuminates the camera. These calibration data are not directly usable for analysis of SGARFACE data, because the flashes are too bright and saturate the FADCs. 

\begin{figure}[!ht]
    \includegraphics[width=0.35 \textwidth,clip]{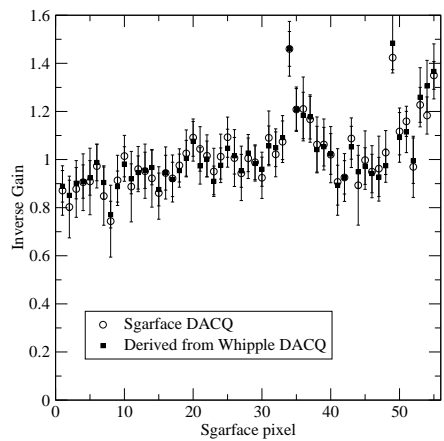}
\caption{ Comparison of the flat-fielding coefficients derived with from SGARFACE and Whipple data. In the case of Whipple data, the individual pixels were remapped to the cluster making up the SGARFACE pixels. The data is from a one minute laser run on MJD 52821, Whipple run gt025056. Error bars show the standard deviation of the distribution. }
\label{fig:gain_laser}
\end{figure}

\begin{figure}[!ht]
    \includegraphics[width=0.35 \textwidth,clip]{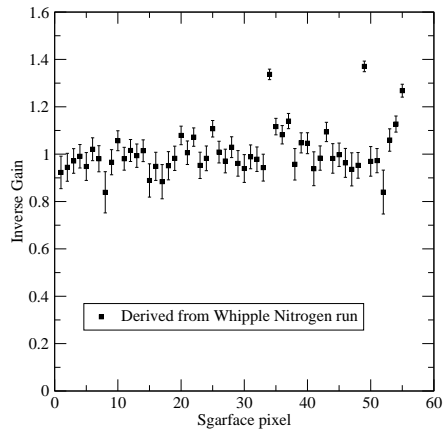}
\caption{ Flat-fielding coefficients shown for SGARFACE channels, but derived from an analysis of the Nitrogen flasher data taken with the Whipple 10~m DAQ system. }
\label{fig:gain_nitrogen}
\end{figure}

The gain uniformity of SGARFACE channels was tested with two methods:

1) Dedicated flat-fielding data were taken at variable brightness levels using a laser system. The laser flashes were recorded by both, SGARFACE and the Whipple 10~m DAQ. The flat-fielding coefficients were determined independently for each system and are shown in Fig.~\ref{fig:gain_laser}, where the gains of the Whipple 10~m have been mapped onto the appropriate SGARFACE channels. Both sets of flat-fielding coefficients are in agreement with each other. The $\chi^2$-difference between the two distribution, calculated using the sample standard deviation of Whipple gain and the standard deviation of the mean SGARFACE gain, is 15.7 with 55 degrees of freedom. Therefore, the flat-fielding coefficients from a Whipple Nitrogen run can be used to predict the true flat-fielding coefficients of the SGARFACE pixel map. Figure \ref{fig:gain_nitrogen}.

\begin{figure}[!ht]
    \includegraphics[width=0.3 \textwidth,viewport= 90 10 550 520,clip,angle=-90]{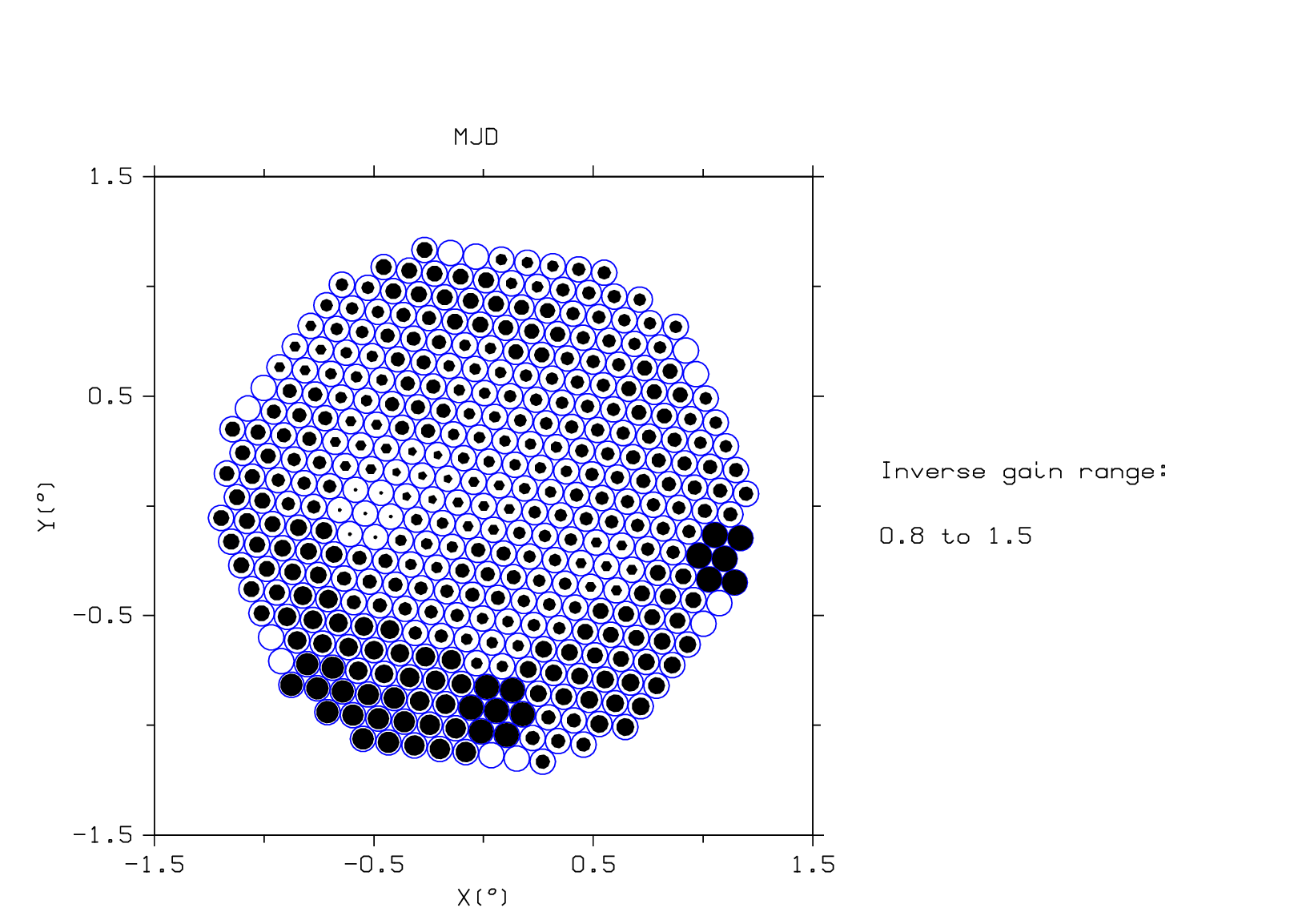}
\caption{ Map of the Whipple camera showing the flat-fielding coefficients taken with a mis-aligned laser. The coefficients are represented by the filled pixels with range between 0.8 and 1.5.}
\label{fig:gain_laser_map}
\end{figure}

The variation of the gain seen in Fig.\ref{fig:gain_laser} is shown in Fig.~\ref{fig:gain_laser_map} in camera coordinates. As can be seen, during these flat-fielding runs the laser was not well aligned with the optical axis of the telescope.

2) The Whipple 10~m flat-fielding coefficients as determined from the Nitrogen-flasher data are shown in Fig.~\ref{fig:gain_nitrogen}. It has been established during the many years of operation of the 10~m, that the Nitrogen flasher uniformly illuminates the field of view, so that any measured gain differences are intrinsic to the camera. The distribution is uniform, except for two outliers, corresponding to SGARFACE pixels 34 and 49. In both cases, the Whipple analysis reveals a low-gain PMT: PMT number 303 corresponding to SGARFACE pixel 49 and PMT number 316 for pixel 34.

Burst of $\gamma$ rays produce broad images of Cherenkov light that is spread over a large number of SGARFACE pixels. Typically, 10 to 30 SGARFACE pixels would contain light, with the gain for SGARFACE pixels varying randomly between 0.8 and 1.2, see Fig.~\ref{fig:gain_nitrogen}. With gains determined to within 20\%, the parameter calculation is effected very little since a large number of pixels are involved. For example, the total amount of detected light for a cluster of 15 pixels has an uncertainty of $0.2/\sqrt{15}=0.05$.
%
%
%
%%
%\bibliographystyle{elsart-num}
%\bibliography{references_Sgarface}
%%

%
%
\end{document}